\documentclass[fleqn,usenatbib]{mnras}

\usepackage{graphicx}	
\usepackage{amsmath}	
\usepackage{amssymb}	
\usepackage{multicol}        
\usepackage{bm}		
\usepackage{pdflscape}	
\usepackage{hyperref}
\usepackage{url}

\newcommand{\kms}{\,km\,s$^{-1}$} 

\newcommand{\angstrom}{\mbox{\normalfont\AA}}

\usepackage[T1]{fontenc}
\usepackage{ae,aecompl}
\usepackage{acro}

\usepackage{newtxtext,newtxmath}

\def\fe{{\sc{Fe}}\/}
\def\fe6087{{\sc [Fe vii]}$\lambda$6087\/}
\def\feii{{Fe\sc{ii}}\/}
\def\ha{{\sc H}$\alpha$}
\def\hb{{\sc{H}}$\beta$\/}

\def\heiiopt{{{\sc H}e{\sc ii}}$\lambda$4686\/}

\def\kms{km~s$^{-1}$}

\def\ni{{[N\sc{i}]}$\lambda$5200\/}
\def\niill{{[N\sc{ii}]}$\lambda\lambda$6548,6584\/}

\def\niib{{[N\sc{ii}]}$\lambda$6584\/}

\def\oiiiopt{{\sc{[Oiii]}}$\lambda\lambda$4959,5007\/}
\def\oiiia{{\sc [Oiii]}$\lambda$5007}

\def\oill{{[O\sc{i}]}$\lambda\lambda$6300,6364\/}

\def\siill{{[S\sc{ii}]}$\lambda\lambda$6716,6731\/}

\usepackage{color,ulem}  
\definecolor{deepblue}{rgb}{0,0,0.5}  
\definecolor{deepred}{rgb}{0.6,0,0}   
\definecolor{deepgreen}{rgb}{0,0.5,0} 
\definecolor{darkgreen}{rgb}{0,0.6,0} 

\DeclareAcronym{MCS}{
  short = MCS,
  long = Megain Courbin Sohy,
  cite = {Magain+98}
}

\title[AGN-Host decomposition]{An iterative method to deblend AGN-Host contributions for Integral Field spectroscopic observations}

\author[Ibarra-Medel et al.]{
H.~Ibarra-Medel$^{1},$\thanks{E-mail: hibarram@astro.unam.mx (HIM)}
C.~ A.~ Negrete$^{2}$,
I. ~ Lacerna$^{3,4}$,  
H.~ M.~ Hern\'andez-Toledo$^{1}$, \and
E.~ Cortes-Su\'arez$^{5}$,
S. F. ~ S\'anchez$^{6}$.
\\
$^{1}$Instituto de Astronom\'{\i}a, Universidad Nacional Aut\'onoma de M\'exico, A.P. 70-264, 04510 CDMX, M\'exico \\
$^{2}$CONAHCyT Research Fellow-Instituto de Astronom\'{\i}a, Universidad Nacional Aut\'onoma de M\'exico, A.P. 70-264, 04510 CDMX, M\'exico \\
$^{3}$Instituto de Astronom\'{\i}a y Ciencias Planetarias, Universidad de Atacama, Copayapu 485, Copiapó, Chile\\
$^{4}$Millennium Institute of Astrophysics (MAS), Nuncio Monseñor Sótero Sanz 100, Providencia, Santiago, Chile\\
$^{5}$Instituto de Astrof\'isica \'Optica y Electr\'onica, Adress: Luis Enrique Erro 1, Tonantzintla, Puebla 72840 M\'exico \\
$^{6}$Instituto de Astronom\'{\i}a, Universidad Nacional Aut\'onoma de M\'exico, A.P. 106, 22800 Ensenada Baja California, México
}

\begin{document}

\label{firstpage}
\pagerange{\pageref{firstpage}--\pageref{lastpage}}
\maketitle

\begin{abstract}
We present a new iterative deblending method to separate the host galaxy (HG) and their Active Galactic Nuclei (AGN) emission with the use of Integral Field spectroscopic (IFS) data. The method decomposes the resolved HG emission from the unresolved AGN emission by modelling the two-dimensional surface brightness (SB) profile of the point-spread function (PSF) and the two-dimensional SB HG continuum simultaneously per each monochromatic slide.  Our method does not require any prior information about the observed SB profile or a detailed fitting of the PSF, making it ideal for the automatic analysis of large galaxy samples. In this work, we test the quality of our method, its advantages, and its disadvantages. We test our method by using a set of IFS mock data cubes to quantify the reliability of our deblending process and further compare our method with the {\sc QDeblend3D} analysis tool. Furthermore, we applied our method to three data cubes selected from the MaNGA survey according to the dominance of either its HG or its AGN. We show that our deblending method is capable of disengaging the bright, nonresolved AGN emission from the HG continuum and its narrow emission lines. However, the decoupling depends on how well the IFS spatially resolves the PSF, and on the relative flux intensity of the HG-AGN. Therefore, the method is ideal for disentangling the bright-flux contribution from AGN-dominated spectra.
\end{abstract}


\section{Introduction}
Active Galactic Nuclei (AGN) manifest in very different ways in the Universe, from those that hide within seemingly normal galaxies to those whose vast amounts of energy are known to far outshine the light of the stars in their host galaxies (HG). This situation immediately leads to questions about the AGN link with the HG and how the AGN energy feedback impacts their evolution. For example, the energy of radio jets emerging from the AGN can heat the interstellar medium and quench the formation of stars in the HG \citep[e.g.,][]{McNamara+00,Bruggen+02,Forman+07,Barai+16,Gaspari+20,Husemann+22}. Furthermore, the intense energy of the AGN can ionise the surroundings of the HG interstellar medium up to kiloparsec scales and form the extended emission line regions and the extended narrow line regions \citep[EELR,ENLR,][]{Stockton+83,Unger+87,Heckman+91,Garcia-Lorenzo+2005,Bolton+07,Sanchez+2007c,Eilers+17,Villar-Martin+18,Chen+19,Husemann+22}. However, it is not completely clear how these processes interact in the complex mechanisms governing the evolution of the HG. For example, how the galaxy environment connects to their nuclear activity, or how the dynamical times of the galaxy are connected with the short and intense energy outburst cycles of their AGN are open questions \citep[e.g.,][]{Peng+10,Hickox+14}. Therefore, studying the HG, their evolution, and their properties is important, particularly during the most energetic phases of their AGN \citep{Lammers+23}. 




The luminosity of an AGN can easily outshine the emission of its HG, depending on the type of AGN \citep{Antonucci+93,Villarroel+14, Cortes+22, Jalan+23}. Consequently, it is essential to distinguish the HG emission from the intense AGN contribution. A major obstacle in image and spectroscopic studies is the high contrast between the AGN emission and its HG, which is usually two to three times fainter than the AGN flux. To address this issue, several approaches based on AGN/HG spatial deconvolution, decomposition, and deblending have been developed from imaging and spectra data. These approaches emphasise the importance of having reliable decomposition methods to separate the spectrum of the central AGN from its HG as much as possible. One of these approaches was developed by \citet[][]{Magain+98}, who implemented a robust deconvolution algorithm that avoids deconvolution artefacts and can be used to retrieve foreground galaxy lensing maps from the residuals \citep[e.g.,][]{Richardson+72,Lucy+74,Skilling+84,Courbin+97,Courbin+98,Burud+98}. Furthermore, \citet{Courbin+2000} presented the spectroscopic version of the \citet[][]{Magain+98} algorithm. They demonstrated that clean HG spectra of objects with severely mixed AGN-HG objects can be accurately extracted.

Many observational studies have used visible imaging to investigate the EELRs of the HG \citep{Kristian+73,Gehren+84,Malkan+84,MMG+84,Smith+86,Hutchings+87,Romanishin+89,Veron-Cetty+90,McLeod+94}. In addition, several works have tried to study the AGN and their HG galaxies properties by decoupling the AGN emission from their HG \citep[e.g.,][]{Bennert+2011,Bennert+2021,Kim+2008,Bruce+2016,Kim+2017,Kim+2019,Li+2021,Li+2023}. They model the surface brightness (SB) profiles of the AGN non-resolved emission and the extended HG emission using the tool {\sc galfit} \citep{Peng02}. Those works show that the HG can be modelled with a \citet{Sersic1963} profile, exploring the implication of 2D modelling of the bulge and disk with different S\'ersic 
indexes. However, all these works indicate the importance of an accurate model of the SB profile of the AGN emission as a point-spread function (PSF). For example, with the Hubble Space Telescope (HST) capabilities, \citet{Li+2021} use the foreground stars to obtain an accurate PSF model to subtract the AGN contribution from the HG. Therefore, for the AGN-HG decomposition, the accurate 2D modelling of the SB significantly impacts the decomposition's success.

During the last decade, Integral Field Spectroscopy (IFS) has significantly improved the way of studying galaxy properties. The IFS provides a way to spatially resolve the spectral properties of galaxies: the resolved ionised gas, the kinematics, and its stellar properties \citep[e.g.,][]{Cappellari+2011,Perez+2013,Ibarra-Medel+16,Tissera+16,Cano-Diaz+16,Cano-Diaz+22,Gonzalez-Delgado+15,Gonzalez-Delgado+2016,Gonzalez-Delgado+2017,Lopez-Fernandez+18,Sanchez+2020,Aquino+2020,Artemi+2022,Ibarra-Medel+22b}. Furthermore, the use of IFS in astronomy has provided the advent of large IFS surveys of galaxies such as the Calar Alto Legacy Integral Field Area survey \citep[CALIFA,][]{Sanchez+2012}, the Sydney-AAO Multi-object Integral field spectrograph survey \citep[SAMI,][]{Croom+2012}, the Mapping Nearby Galaxies at the Apache Point Observatory survey \citep[MANGA,][]{Bundy+2015}, which is part of the Sloan Digital Sky Survey phase IV \citep[SDSS-IV,][]{Abdurro+22}, and more recently the upcoming Local Volume Mapper from SDSS phase V \citep[LVM,][]{Almeida+23}.  Other works explored the point-spread function (PSF) deconvolution to study the kinematic properties of galaxies. In this direction, \citet{Chung+21} presented an alternative PSF deconvolution algorithm based on the Lucy-Richardson algorithm \citep{Lucy+74,Richardson+72}. They used IFS data from MaNGA and generated a set of IFS mock observations to demonstrate that the algorithm can recover the true stellar kinematics. Consequently, the IFS offers an unparalleled opportunity to analyse the spatially resolved characteristics of the ionised gas and the HG's extended emission-line regions (EELRs).

One of the first AGN-deblended approaches with IFS was presented by \citet{Sanchez+2004a} and \citet{Garcia-Lorenzo+2005}, they present an adapted version of {\sc galfit} \citep{Peng02} called {\sc galfit3d} \citep{Sanchez+2006c}. With this tool, they studied the central region of 3C 120 and disentangled the AGN emission using IFS data and HST WFPC imaging. In this study, they separated the AGN emission from the EELR and detected six structures associated with a previous merger event and interactions with the radio jet observed in the central region of 3C 120. However, the implementation of {\sc galfit3d} implies the use of {\sc galfit} to model the HG and the AGN surface brightness in many wavelength slides of the data cube, requiring a field star to model the PSF. Later, \citet{Christensen+06} used IFS data to study the AGN and its HG EELRs. They discussed three techniques for subtracting AGN emissions. The first approach involves using synthetic, adjustable narrow-band filters to observe a set of emission lines and measure the adjacent continuum, thus allowing the nuclear emission to be removed and narrow-band images of the HG EELRs to be produced. The second is an iterative method that assumes that the spatially resolved spectrum of a point source (AGN) is the same across adjacent spaxels but scaled by flux. To begin, they combined the central AGN spectrum within a 2-arcsecond aperture. With this extracted spectrum, they calculated the scale factor of the individual spaxels at the loci of an intense emission line to create a new data cube. This data cube contains the central extracted spectrum but is scaled spaxel by spaxel. Then, they subtracted this data cube from the original data to obtain a residual data cube. This residual data cube is again removed from the original cube to get the AGN data cube \citep{Sanchez+2004a}. This process is repeated three times to obtain a pure EELR data cube. The third method involves implementing a two-dimensional fitting of the PSF to model the AGN spectrum for each spectral sampling or monochromatic slice. This technique was previously implemented by \citet{Wisotzki+03} and later improved by \citep[][]{Sanchez+04,Sanchez+2004a,Sanchez+06,Sanchez+2007c,Mendez-Abreu+2017,Mendez-Abreu+2019,Mehrgan+2019}. The approach they used is based on the assumption that the full-width half maximum (FWHM) of the PSF changes smoothly across the wavelength dispersion. Then, the PSF model is interpolated and subtracted to recover an extended emission-line region (EELR) data cube. \citet{Husemann+08} improved the second methodology proposed by \citet{Christensen+06} to study the EELRs from IFS data. \citet{Husemann+08} use the broad emission lines within the spectral region of H$\alpha$ and H$\beta$ to obtain the appropriate spatially resolved scale factor of the point source flux of the AGN and construct an AGN cube. Therefore, they used the AGN cube to recover a data cube from the EELR. Later, \citet{Husemann+13} continued to improve and develop their deblending algorithm, and finally they presented it as the {\sc QDeblend3D}\footnote{\url{https://github.com/brandherd/QDeblend3D} and \url{http://www.bhusemann-astro.org/?q=qdeblend3d}} software tool \citep{Husemann+14}.  

More recently, \citet{Husemann+22} improved their deblending algorithm to implement a wavelength dependency on the PSF. They interpolate the PSF variations from specific spectral windows. However, {\sc QDeblend3D} still needs prior 
information of the HG/AGN relative fluxes to perform the PSF model to subtract and remove the AGN contribution from the HG emission. Hence, a wrong HG/AGN relative flux value will return to a bad AGN/HG decomposition, implying the need for a previous 2D surface brightness modeling using external tools such as {\sc galfit}. On the other hand, \citet{Mendez-Abreu+2019} and in their successive articles apply the 2D surface brightness modeling to spectroscopically disentangle the structural properties (bar, disk and bulge) of galaxies observed with IFS. However, the \citet{Mendez-Abreu+2019} methodology is not intended to perform an AGN/HG decoupling but to spectroscopically decouple the bar, disk and bulge. With the advent of the James Web Space Telescope (JWST), several groups have been working on a method to provide a way to decouple the non-resolved AGN emission from the resolved HG emission using the IFS data of the JWST \citep[e.g.,][the Q3D group]{Wylezalek+2022, Veilleux+2023,Vayner+2023}. They created the tool named {\sc q3dfit}\footnote{\url{https://q3d.github.io/}} that, in a similar way as {\sc galfit3d}, it requires to previously have an accurate model of the JWST PSF to subtract the AGN emission from the HG emission.

We present a new implementation of AGN-HG decomposition. The presented method is a new iterative method designed for IFS data that models the two-dimensional AGN point source and the two-dimensional HG continuum at the same time in each monochromatic slide. This method focuses on AGN-dominated spectra and aims to decouple the resolved HG spectra that are completely buried on the strong non-resolved AGN emission. The principal difference from previous methods is that we designed our methodology to not depend on prior information of the 2D surface brightness PSF and the HG profiles. Our methodology does not need to model a foreground star or the HG previously to get the PSF profiles. In addition, our method does not require interpolating the PSF between different spectral regions.

This paper is organized as follows. In Section \ref{sec:method}, we describe, in a detailed form, the workflow of our deblending method, the inputs and its outputs. In Section \ref{sec:mocks}, we test the precision of our method by implementing a set of IFS mocks that simulate the resolved emission of a set of AGN. In Section \ref{sec:realdata}, we implement our method using three IFS MaNGA galaxies. In that section, we disentangle the AGN/HG spectra, analyse the emission lines of the deblended AGN/HG spectra, and present a characterisation of their nuclear activity and its black-hole masses. Finally, in Section \ref{sec:summary}, we review the advantages and drawbacks of the method.

\begin{figure*}
\includegraphics[width=2\columnwidth]{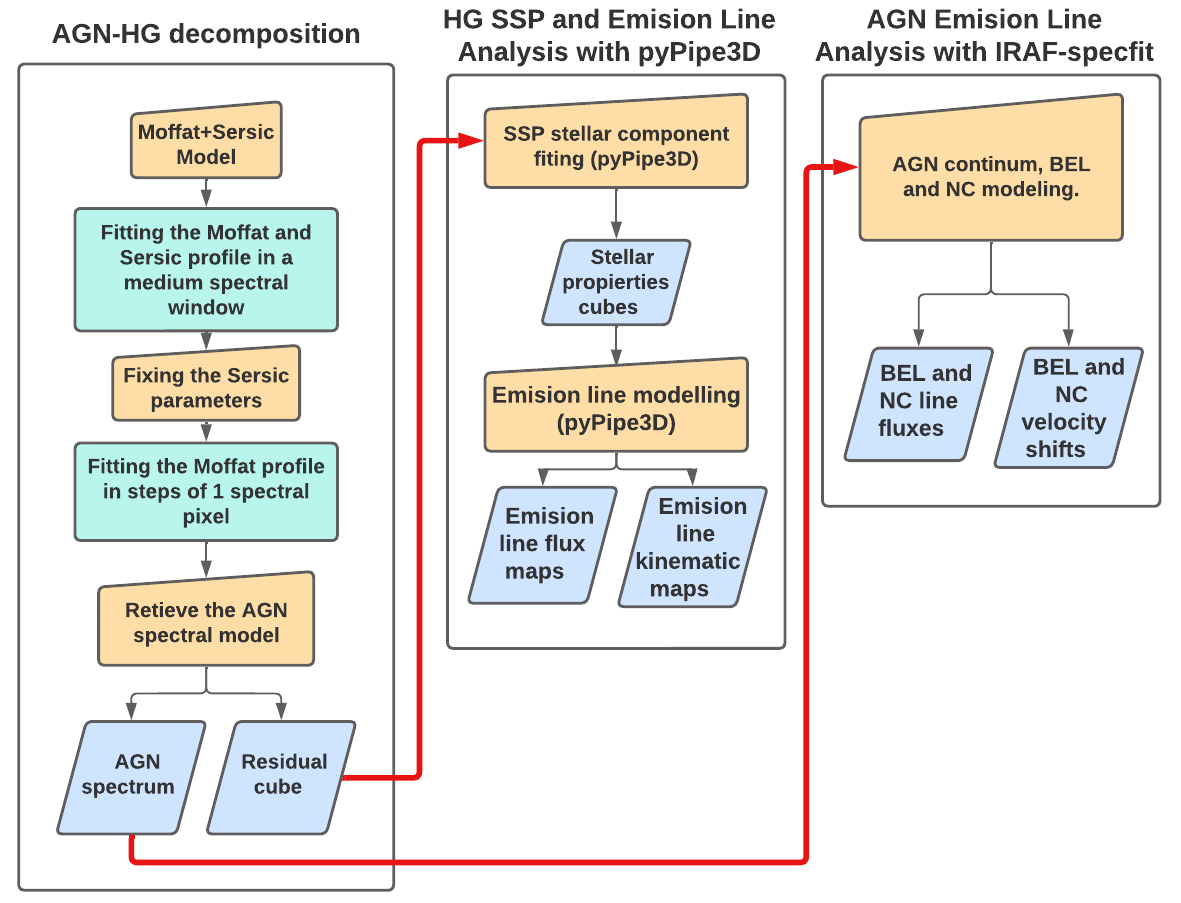}
\caption{Workflow diagram of our iterative AGN-HG deblending method. The boxes indicate the main process of the method: The AGN-HG decomposition. In addition, we add an optional process for the HG and AGN spectral analysis: The HG single stellar population and emission line analysts done by {\sc pyPipe3D}, and the AGN analysis of the broad emission lines (BEL) and the narrow components (NC) done with {\sc IRAF-specfit}.}
\label{fig:diagram}
\end{figure*}

\section{The Method}\label{sec:method}
The methodology underlying this article has been implemented in a code written in python3 that is publicly available in a GitHub repository called {\sc AGNdecompose}\footnote{\url{https://github.com/hjibarram/AGN_decompose}}. The method uses a 2D surface brightness decomposition to extract the AGN spectrum from the HG spectrum. The decomposition models the SB of the observed galaxy as the contribution of two profiles: a \citet{Sersic1963} profile for an extended contribution (HG) and a \citet{Moffat+1969} profile for the unresolved contribution (AGN). The works of \citet{Kim+2008} 
demonstrated that using a single Sersic profile is sufficient to decompose the AGN emission from the host. However, we plan for future development to include the \citet{Mendez-Abreu+2019} methodology to include more complex modelling of the 2D surface brightness of the HG. The methodology workflow for our 2D AGN-HG decomposition is illustrated in Figure \ref{fig:diagram}. Our method works in a similar fashion to {\sc galfit3d} \citep{Garcia-Lorenzo+2005} but with the difference that we consider a PSF that varies with the wavelength, as we describe below. Also, our method does not require an accurate model of the PSF with a foreground star as it is needed for {\sc galfit3d} and {\sc q3dfit}. This method has the advantage of being applied in an automatic form without the need for external feedback and any additional modelling. Hence, one of the main objectives of this work is to explore and quantify the inaccuracies in the AGN-HG deblending by characterizing the PSF with only the AGN emission on IFS observations.

\subsection{Our method: The AGN-HG spectra decomposition.}

We model the non-resolved AGN contribution as a point source with a PSF SB profile that follows a Moffat \citep{Moffat+1969,Trujillo+2001} function with the following form:

\begin{equation}
    F_{psf}(i,j)=A_t\times\left(1+\frac{R(i,j)^2}{\alpha^2}\right)^{-\beta}.\label{EQ1}
\end{equation}
In this case, $A_t$ is the peak value of the PSF profile, $x_0$ and $y_0$ are the central positions of the profile, $\alpha$ is the dispersion and $\beta$ traces the size of the extended wings of the PSF. Mathematically, as $\beta$ increases, the profile asymptotically tends to be a Gaussian profile \citep{Trujillo+2001}. The value of $R(i,j)$ is the distance from the PSF centered to its isophotal value. For an ideal case, the isophotal can have a circular shape; however, our model allows us to define an elliptical shape by defining $R(I,j)$ as:
\begin{equation}
\begin{split}\mathbf{
    R(i,j)=\sqrt{\frac{(1-e^2\sin^2[\arctan(\Delta_x,\Delta_y)-\theta])\times(\Delta_x^2+\Delta_y^2)}{1-e^2}}}\\
    \mathbf{\Delta_x=i-x_0}\\
    \mathbf{\Delta_y=j-y_0}\\
    \label{Eq:Radi}
\end{split}
\end{equation}

Where $e$ and $\theta$ are the PSF isofotal's ellipticity and positional angle. If $e$ is  zero, i.e., a circular isophotal shape, the Equation \ref{Eq:Radi} returns to the standard definition of a circular radius: $R(i,j)=\sqrt{(i-x_0)^2+(j-y_0)^2}$. If $e$ is not zero, $R(i,j)$ returns the mayor axis of the elliptical isophotal. The method has the option to fix $e$ and $\theta$, or level them as an extra of two variables to fit.

For the extended spatially resolved source, we use the classical definition of a \citet{Sersic1963} profile defined as:

\begin{equation}
\begin{split}
   F_{ext}(i,j)=I_o\times  exp \left\{  -b_n\left[ \left(\frac{R'(i,j)}{R_e}\right)^{1/n_s}-1\right] \right\}\\ 
   R'(i,j)=\sqrt{(i-x_0)^2+(j-y_0)^2}.
\end{split}
\end{equation}

Where $I_o$ and $R_e$ are the values of the half-light intensity and radius, and $n$ is the S\'ersic index. In this case, for simplicity, we assume a circular isophotal shape. For the aim of this method, a perfect model of the extended emission is not needed; we require an educated guess of the extended background to fit the intense central emission of the AGNs. However, in a future release of the method, we plan to implement more complex models, including disks, bars and asymmetries.

Therefore, the model fits the observed SB profile as $F_{psf}+F_{ext}$. We set the central positions $x_0$ and $y_0$ as the same positions for $F_{psf}$ and $F_{ext}$. The total number of free parameters for the non-constrain mode (see below) are nine (eleven): $At$, $x_0$, $y_0$, $\alpha$, $\beta$, $I_e$, $b_n$, $R_e$, and $n_s$ ($\mathbf{e}$, $\mathbf{\theta}$). To obtain the best value for the free parameters, we use the \citet{emcee} implementation of the Affine Invariant Markov chain Monte Carlo (MCMC) Ensemble Sampler with the use of the python package {\sc emcee}\footnote{https://emcee.readthedocs.io/en/stable/}. 

With the SB modelling, the AGN/HG decomposition works as follows (see left panel of Figure~\ref{fig:diagram}):


\begin{itemize}
\item[a)] SB fit within spectral window slides: We divide the spectral range into a set of spectral slides of ten spectral pixels each. For each slide, we integrate the spectral information to boost the SNR SB map by $\sqrt{10}$ times. For the case of MaNGA, this is equivalent to a window of 15 \AA\ since MaNGA has a spectral sampling of 1.5 \AA\ per spectral pixel. Then, we perform a non-constraint 2D SB modelling of $F_{psf}+F_{ext}$ for each slide and obtain the best-fit values of $A_t$, $x_0$, $y_0$, $\alpha$, $\beta$, $I_o$, $b_n$, $R_e$, and $n_s$ ($\mathbf{e}$, $\mathbf{\theta}$). Therefore, we obtain a low-resolution spectral dependence for each of the nine (eleven) variables. 

\item[b)] Fixing the S\'ersic parameters: From the previous step, we obtain their low-resolution smoothing spectral response for only the $b_n$, $n_s$, and $\beta$. The method first smooths each spectral response function with a Gaussian kernel of the size of two times the spectral window used in step a). Then, the method applies a linear interpolation of the values at the full spectral sampling. If the PSF isophotal is not circular, the variables $e$ and $\theta$ are also interpolated to the full spectral range. The method exploits the assumption that the host S\'ersic parameters $b_n$, $n_s$, and the PSF parameters $\beta$ ($e$, $\theta$) cannot change drastically as a function of wavelength. In other words, those parameters can only monotonically change as a function of wavelength.

\item[c)] SB fit for all spectral sampling: Once we obtain the interpolated values of $b_n$, $n_s$, and $\beta$ ($e$, $\theta$), we proceed to re-fit the SB profile but now at each spectral sampling or monochromatic slide (one spectral pixel). We are able to reduce our space parameter from nine (eleven) to six: $A_t$, $x_0$, $y_0$, $\alpha$, $I_o$, and $R_e$. The code has a set of keywords to interpolate $R_e$, $I_0$, $x_0$, and $y_0$ values, leaving only  $A_t$ and $\alpha$ as free parameters. However, those keywords are not activated as default values.

\item[d)] Reconstruction of the nonresolved point source (AGN) spectral model: We reconstruct the point source spectra using the full spectral responses of $A_t$, $x_0$, $y_0$, $\alpha$, and $\beta$ ($\mathbf{e}$, $\mathbf{\theta}$). We use Equation~\ref{EQ1} to recover the flux map at each spectral pixel and then reconstruct the AGN spectral cube. The parameters $I_o$ and $R_e$ are the only free parameters to model the HG S\'ersic profile.

\item[e)] Retrieve the AGN and HG spectra: With the AGN model cube, we simply obtain the residual cube (Original Cube minus AGN Cube). The residual cube will contain the extended emission associated with the HG that includes the HG EELR. On the other hand, the AGN cube will contain the point source unresolved emission: the AGN spectra including the Broad Emission Lines (BELs), their Narrow Components (NC), and the nonthermal continuum. 
\end{itemize}

We finally obtain the spectral responses of the nonresolved emission: $A_t(\lambda)$, $\alpha(\lambda)$, $\beta(\lambda)$, $x_0(\lambda)$, $y_0(\lambda)$, and  $e(\lambda)$, $\theta(\lambda)$ if the PSF elliptical isofotal is activated. With this information, we proceed to extract the pure unresolved (AGN) emission flux with the next analytical expression:

\begin{equation}
    \begin{split}
        F_{AGN}(\lambda)=\int F_{psf}(r,\lambda)dr=\frac{\pi \alpha(\lambda) ^2 A_t(\lambda)}{\beta(\lambda)-1}.
    \end{split}
    \label{eq-fluxtot}
\end{equation}

This expression comes from the integral of Equation \ref{EQ1} or the total flux of the PSF 2D SB profile at each spectral sampling. Therefore, the method is able to obtain the total non-resolved flux spectrum. In addition, the method produces an AGN-free HG cube. The user can apply diverse analysis tools to study the HG AGN-free properties and the AGN HG-free spectra. This paper implements the {\sc pyPipe3D} and the {\sc specfit} (IRAF's task)-specific tools to retrieve the stellar, the HG EELR, and the AGN BEL, NC and the nonthermal continuum proprieties.
This marks the end of our AGN-HG decomposition methodology. 
Next, we describe the methodologies used to analyze the disentangling HG and the AGN spectra, remarking that those methods are no longer part of our AGN/HG decomposition method. However, in this work, we use {\sc pyPipe3D} and {\sc specfit} to quantify the accuracy of our blending method.

\subsection{The HG SSP and emission line analysis.}

For the HG analysis, we use the {\sc pyPipe3D}\footnote{\url{https://gitlab.com/pipe3d/pyPipe3D/}} analysis tools. With the residual data cube, we proceed to model the stellar spectra. We follow the next steps (see the middle panel of Figure \ref{fig:diagram}): 

\begin{itemize}

\item[a)] The single stellar population (SSP) fitting: We use the {\sc pyPipe3D} spectral fitting tool. {\sc pyPipe3D} performs a  full stellar population synthesis \citep[SPS\textcolor{cyan}{,}][]{Lacerda+2022,Sanchez16a,Sanchez16b,Sanchez+2022}, and its workflow is fully described in \citet{Lacerda+2022} and \citet{Sanchez+2022}. It works as follows: It performs a spatial segmentation to achieve an SNR of 50. Then, {\sc pyPipe3D} runs a first model of the stellar spectra per each spatial bin and obtains a gas spectrum from which it fits the strong emission lines. Then {\sc pyPipe3D} runs an SSP analysis using a full stellar library. In this case, we use the MaSTAR-sLOG stellar library defined in \citet{Sanchez+2022}. It uses 273 SSP templates that cover 39 stellar ages (from 1 Myr to 13.5 Gyr) and seventh metallicities ($Z_{\star} =$, 0.0001, 0.0005, 0.002, 0.008, 0.017, 0.03, 0.04). This library uses a \citet{Salpeter+55} initial mass function (IMF). In addition, {\sc pyPipe3D} uses a \citet{Cardelli+1989} dust extinction law for internal extinction. Following the SSP fitting, {\sc pyPipe3D} reconstructs a set of 2D maps that contains the properties of the HG stellar population properties. {\sc pyPipe3D} creates 2D maps with the same number of spaxels of the original datacube by undoing the initial segmentation binning \citep{Lacerda+2022}.



\item[b)] Finally, {\sc pyPipe3D} obtains a gas cube by subtracting the stellar model cube from the HG residual cube. The gas cube will contain the total emission of the HG pure EELR. The emission line fitting is fully described in \citet{Lacerda+2022} and \citet{Sanchez+2022}, but we summarize the process. With the gas cube, {\sc pyPipe3D} implements a spaxel by spaxel fit of the emission lines. {\sc pyPipe3D} models the emission lines using a Gaussian profile to retrieve the amplitude, velocity dispersion and velocity shifts of a set of 50 strong and weak lines. This set includes the \ha,\ \hb,\ \niill\ and \oiiiopt\ lines. The emission line fit is done following the next workflow: {\sc pyPipe3D} implements a random method (called RND) based on a Monte-Carlo (MC) $\chi^2$ minimization to retrieve the best model parameters (amplitude, velocity dispersion and the velocity shifts). Once this step is done, {\sc pyPipe3D} uses the best-fit parameters of the MC as prior value for a Levenberg-–Marquardt method to re-explore the space parameter and find the best model fit of each emission line. Therefore, {\sc pyPipe3D} will return a set of spatially resolved maps of the fluxes, velocity dispersion, velocity shifts, and equivalent widths.

\end{itemize}

\subsection{The AGN spectra analysis.}

To analyze the recovered AGN spectra, we use the {\sc IRAF-specfit} task \citep{kriss94}. {\sc specfit} is an iterative tool within the {\sc IRAF} system that spectroscopically fits various emission and absorption lines, and continuum models at the same time. It works as follows:

\begin{itemize}
    \item[a)] The user must select a combination of functional forms to model all the emission/absorption line components, starting from an underlying continuum constructed by an initial flux guess and a power law index. Then, {\sc specfit} model the strong BELs followed by the NCs and other additional components such as the well-known \feii\ optical pseudo continuum and absorption lines, if needed. The task also permitted us to choose Gaussian or Lorentzian line profiles given as input of their intensity, central wavelength, and FWHM. An ASCII file should be provided with the initial guesses of the selected components. This selection allows us to model complex line systems such as blended emission and absorption lines, absorption edges, and complex continuum features such as the \feii\ emission and extinction. All emission components linearly contribute to modelling the input spectra. 
    
    \item[b)] The fitting is carried out by a $\chi^2$ minimization using a Marquardt, Gridfit or Alternate algorithms taken from the Numerical recipes \citep{Press+1986, kriss94}. The minimization ended when the tolerance requirements met the input specification or reached the maximum number of iterations. Finally, the best-fitting parameters for each spectral component are saved in a log file. Another file for plotting could also be provided with all flux components as functions of the wavelength.
\end{itemize}

\section{Bias and Uncertainties}\label{sec:mocks}

\begin{figure}
\includegraphics[width=\columnwidth]{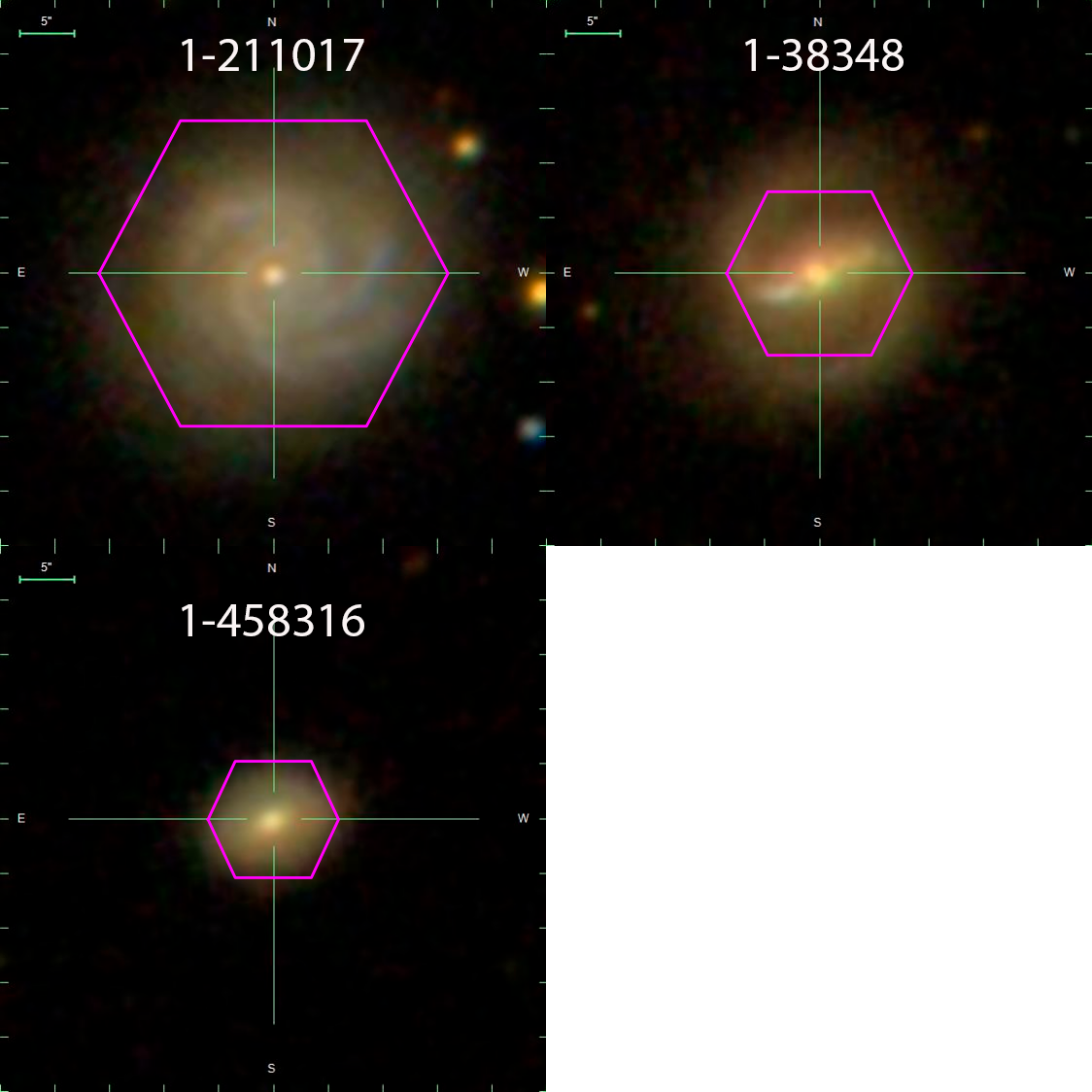}
\caption{Input host galaxy resolved spectra taken from MaNGA. The MaNGA-ID 1-211017 (top left) is used for the high-resolution case, 1-38348 (top right) for the intermediate-resolution case, and 1-458316 (bottom left) for  the low-resolution case.}
\label{fig:MANGA}
\end{figure}

\begin{figure*}
\includegraphics[width=2\columnwidth]{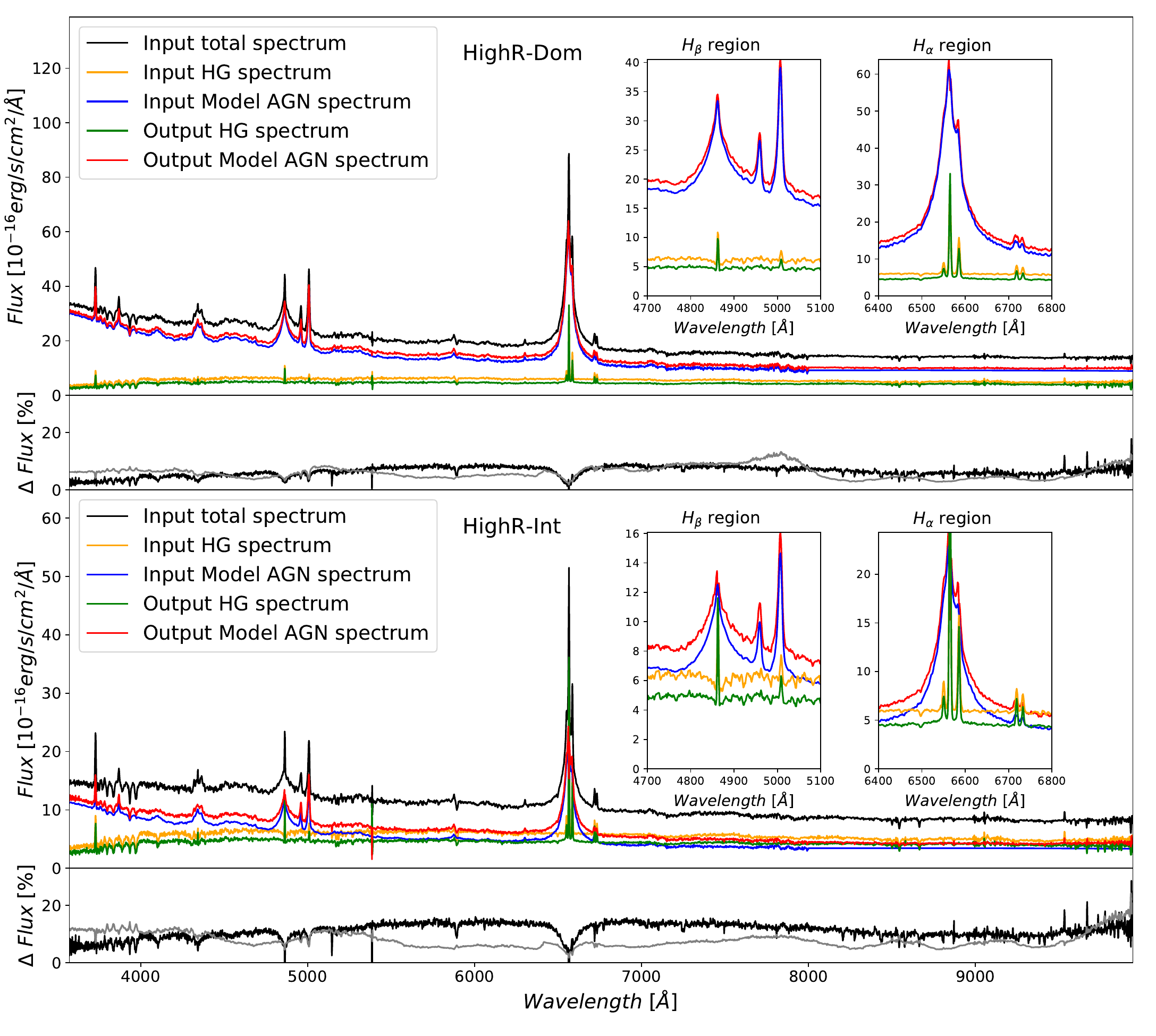}
\caption{High-resolution AGN-dominated (HighR-Dom) mock decomposed spectra (two upper panels), and High-resolution intermediate-AGN (HighR-Int) mock decomposed spectra (two lower panels). For each mock spectra, the main panel shows the total mock spectrum (AGN+HG) as a black solid line, the mock input HG spectrum as a yellow solid line, the input AGN spectrum as a solid blue line, the decomposed output HG spectrum as a green solid line, and the decomposed output AGN spectrum as a red solid line. The two inset plots show a zoom-in of the $H_{\beta}$ and $H_{\alpha}$ regions. The lower panel shows the normalised residual flux defined as $(Flux_{out}-Flux_{in})/Flux_{tot}$ as a solid black line. The solid grey line is the normalised error of the input spectrum defined as Error/$Flux_{tot}$. The spectrum is the total spectrum within 2.5 arcsecs of the centre of the IFU-FoV.}
\label{fig:high_resolution}
\end{figure*}

\begin{figure*}
\includegraphics[width=2\columnwidth]{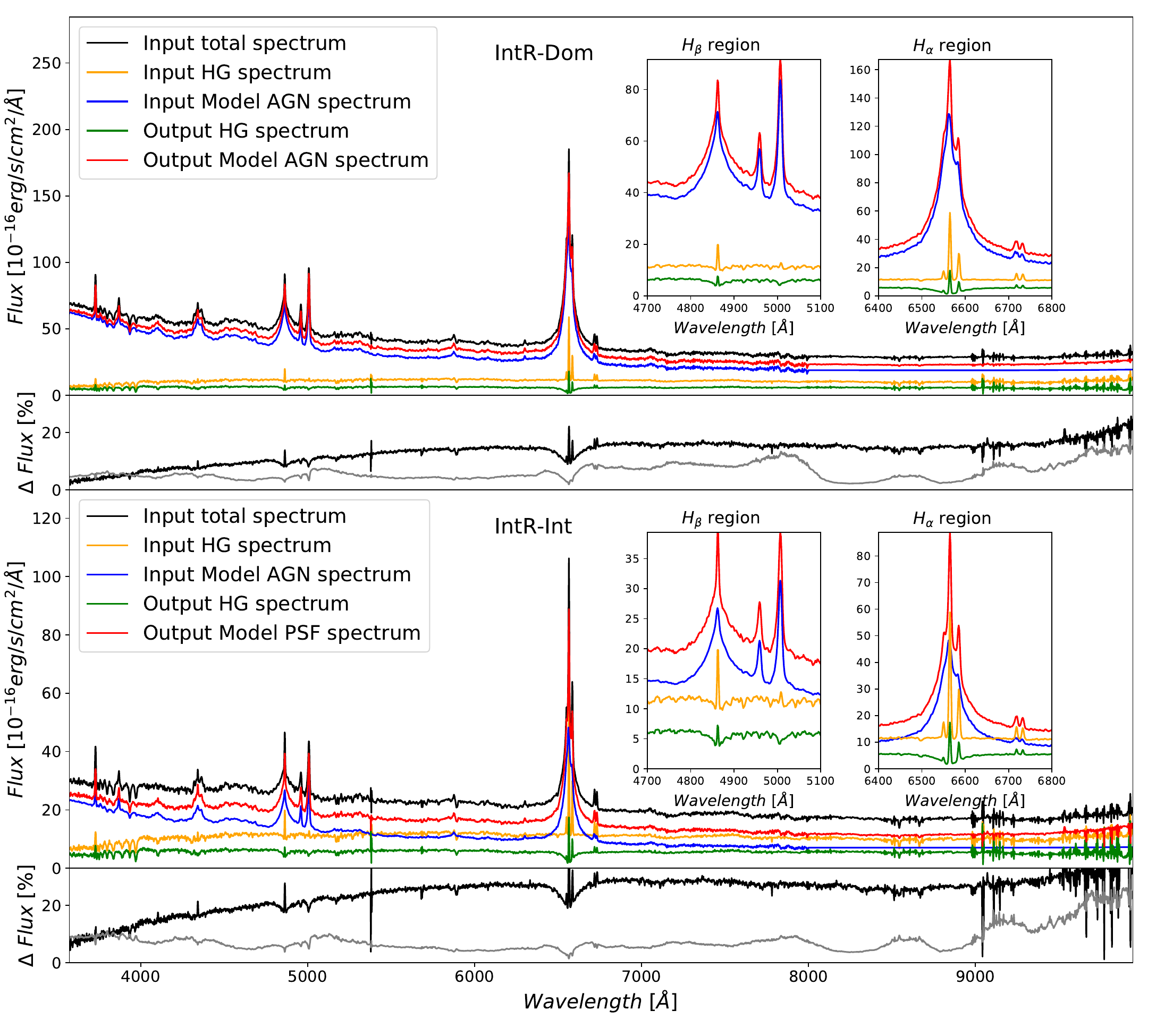}
\caption{Intermediate-resolution AGN-dominated (IntR-Dom) mock decomposed spectra (two upper panels) and the Intermediate-resolution intermediate-AGN (IntR-Int) mock decomposed spectra (two lower panels). The colour code is the same as in Fig. \ref{fig:high_resolution}.}
\label{fig:intermediate_resolution}
\end{figure*}

\begin{figure*}
\includegraphics[width=2\columnwidth]{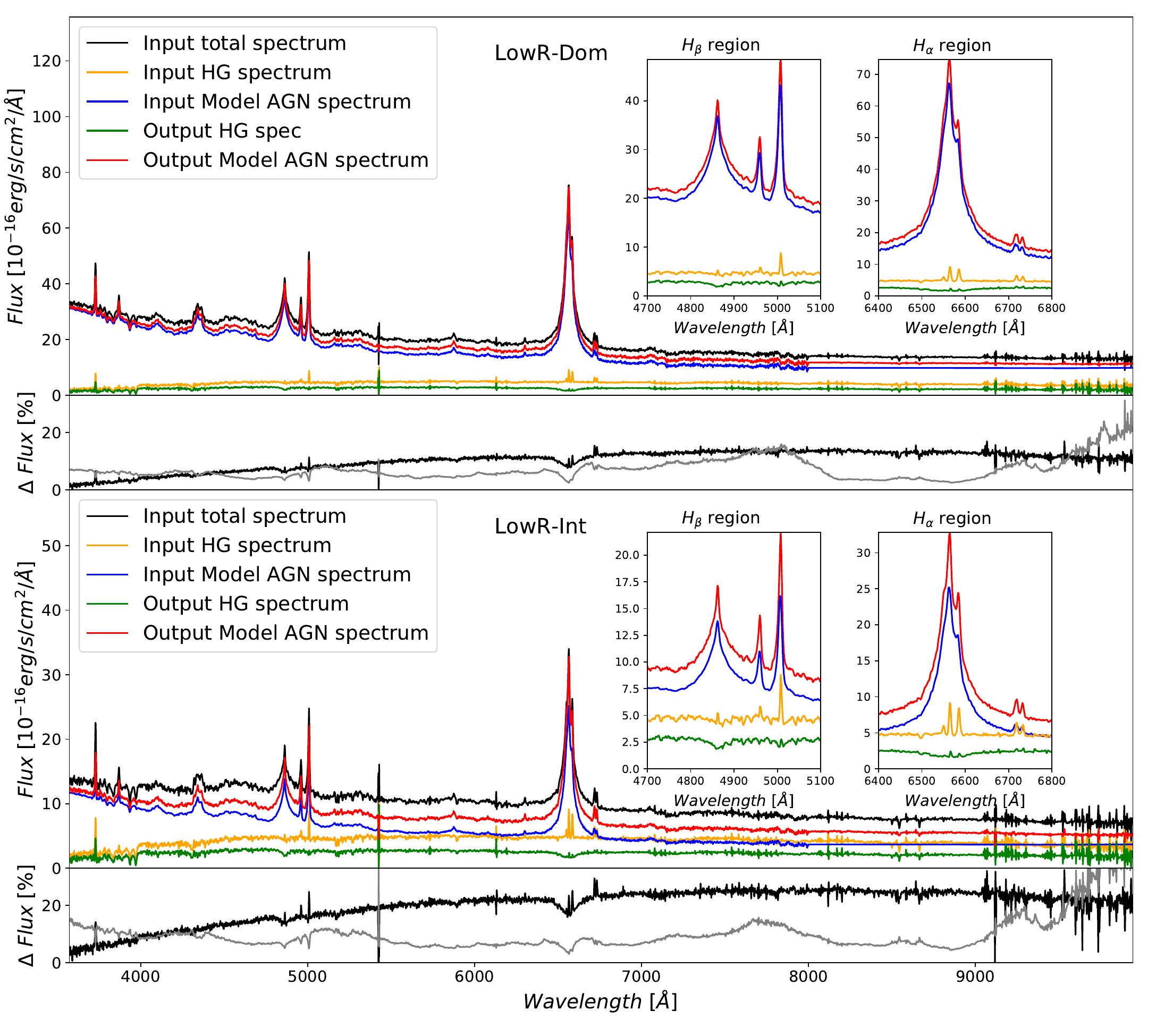}
\caption{Low-resolution AGN-dominated (LowR-Dom) mock decomposed spectra (two upper panels) and the Low-resolution intermediate-AGN (LowR-Int) mock decomposed spectra (two lower panels). The colour code is the same as 
Fig. \ref{fig:high_resolution}.}
\label{fig:low_resolution}
\end{figure*}

\subsection{Mock IFU Spectra}\label{sec:Mock IFU Spectra}

To estimate the bias and uncertainties of our deblending process, we generate a set of six mock IFU observations with mock AGN emission spectra. To generate those mocks, we used a set of three MaNGA \citep{Bundy+2015} galaxies: MaNGA-IDs 1-211017, 1-38348, and 1-458316, see Figure~\ref{fig:MANGA}. The three galaxies have strong EELRs without the presence of any AGN activity. Therefore, the selection of these galaxies is ideal to be the input of HG-resolved spectra for the mocks. The galaxy 1-211017 is a 127 MaNGA IFU fiber bundle (32" in diameter) to test a high spatial resolution IFS observation. The galaxy 1-38348 is a 37 IFU bundle (17" in diameter) to test the intermediate spatial resolution IFS. Finally, the galaxy 1-458316 is a 19 IFU bundle (12" in diameter) to test the low spatial resolution IFS. Assuming that the MaNGA FOV covers at least 1.5 effective radii ($R_e$), the difference among the 127, 37 and 19 bundles implies a spatial resolution of 0.05, 0.09 and 0.13 $R_e$ per spaxel. MaNGA has a spaxel size of 0.5" and a fiber diameter of 2.5" \citep{Law+2016}.

To generate the AGN mock spectra, we first obtain the composite spectra of 1000 luminous quasars from the SDSS-DR16th quasar catalog \citep{SDSSDR17Q}. To construct the composite spectra, we follow the methodology described in \citet{Vanden+2001}. This methodology normalizes the set of quasar spectra at a rest-frame wavelength window that is common among all the spectra and obtains the average. The composite spectra are then redshifted to the rest-frames of the three MaNGA galaxies, and it is used to reproduce the total AGN flux emission ($F_{AGN}$) of the mocks. 

We define the value of the Moffat FWHM PSF size from the analytical description in Equation 1 of \citet{Trujillo+2001}. This analytical description returns the dependence of the PSF size as a function of $\alpha$ and $\beta$. Then, we set a wavelength dependency of the PSF as a monotonic function that changes from 2.5 arcsec at 3,500 $\angstrom$ to 3 arcsec at 11,500 $\angstrom$. Also, we define that $\beta$ should be a constant value of 26 dex. The PSF properties are based on observed MaNGA PSF profiles taken from stars of the MaNGA Stellar library program \citep{Yan+2019}. We fit the 2D PSF SB profiles of the MaNGA stars IFS observations, to obtain a realistic MaNGA PSF profiles as a functions of wavelength. Therefore, we can provide an accurate PSF ($\alpha$) variation based on actual MaNGA IFS data. With the value of the PSF and $\beta$, we can solve the value of $\alpha$ from Equation 1 of \citet{Trujillo+2001}. Then, we use our Equation \ref{eq-fluxtot} to get the value of $A_t$ as a function of $\alpha$ and $\beta$. We define a monotonic shift of the AGN centroid ($x_0, y_0$) from the centre of the MaNGA IFU Field-of-View (FoV) at 3,500 $\angstrom$ to 0.5 arcsec at 11,500 $\angstrom$ in the north to south direction. Finally, with the values of $A_t(\lambda)$, $\beta(\lambda)$, $\alpha(\lambda)$, $x_0(\lambda)$, and $y_0(\lambda)$, we can use our Equation \ref{EQ1} to obtain the 2D AGN spectra within the MaNGA IFU FoV. 

The final IFU mock spectra are the sum of the 2D HG spectra with the 2D AGN spectra. In this step, we rescale the total flux of the 2D AGN spectra to be 1.2 and 3.2 times the total flux of the HG spectra within 5,100 to 5,300 $\angstrom$. These values can represent the intermediate AGN-HG spectra and the AGN-dominated spectra. Therefore, we obtained six mock IFS observations: three spatial-resolution IFS observations (highR, intR, lowR) with intermediate and dominated (Int, Dom) AGN spectra each. 

\subsection{Results of the Mock Decomposition}

\begin{figure*}
\includegraphics[width=2\columnwidth]{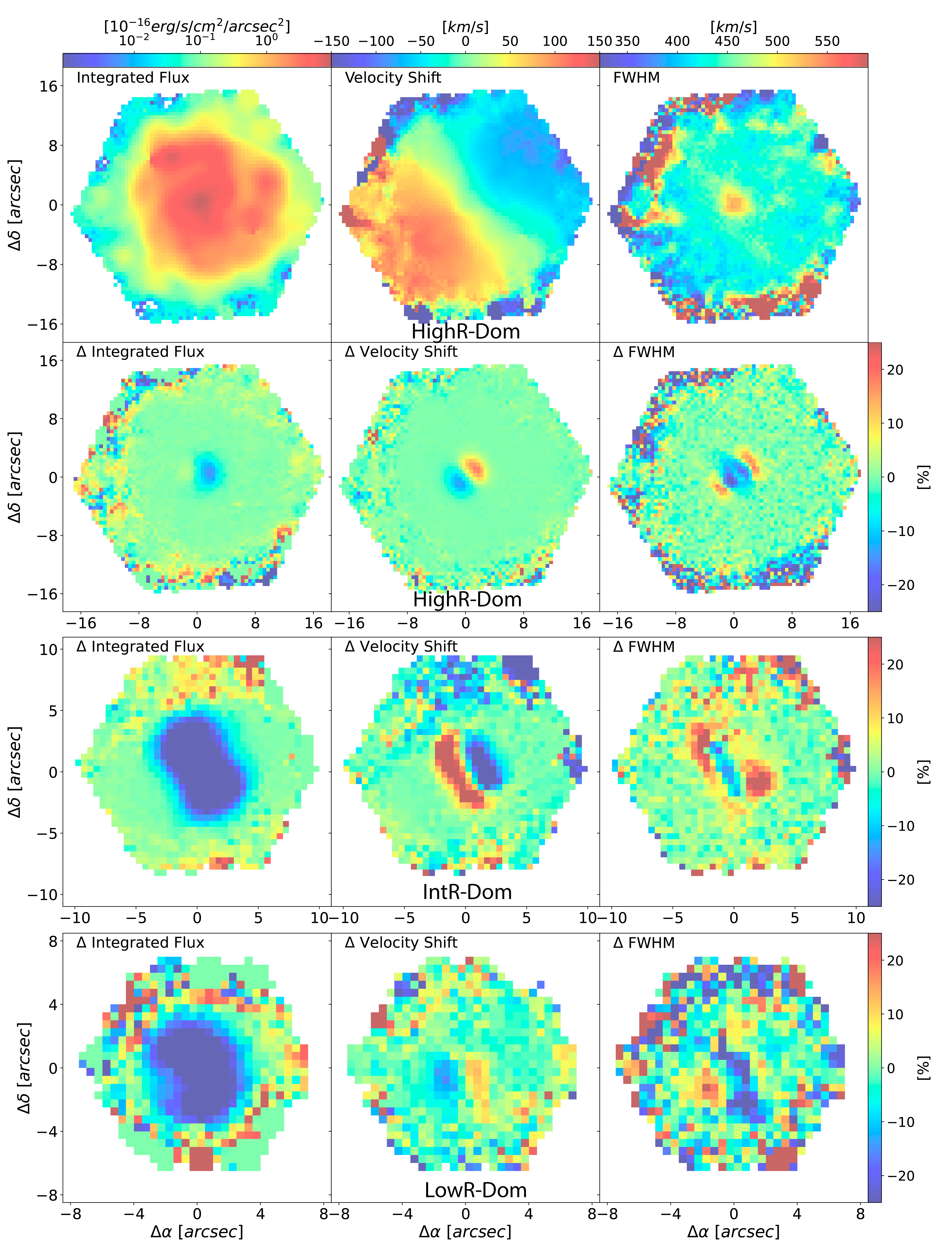}
\caption{The HG \ha\ EELR absolute residual maps for the High (second-row panels) and its input HG values (first-row panels). The absolute residual maps for the Intermediate and Low-resolution mock cases are the third and fourth-row panels. The left, middle and right panels show the spatially resolved absolute residual of the total \ha\ emission flux, velocity shift and FWHM, respectively.}
\label{fig:ELMock}
\end{figure*}

\begin{figure*}
\includegraphics[width=2\columnwidth]{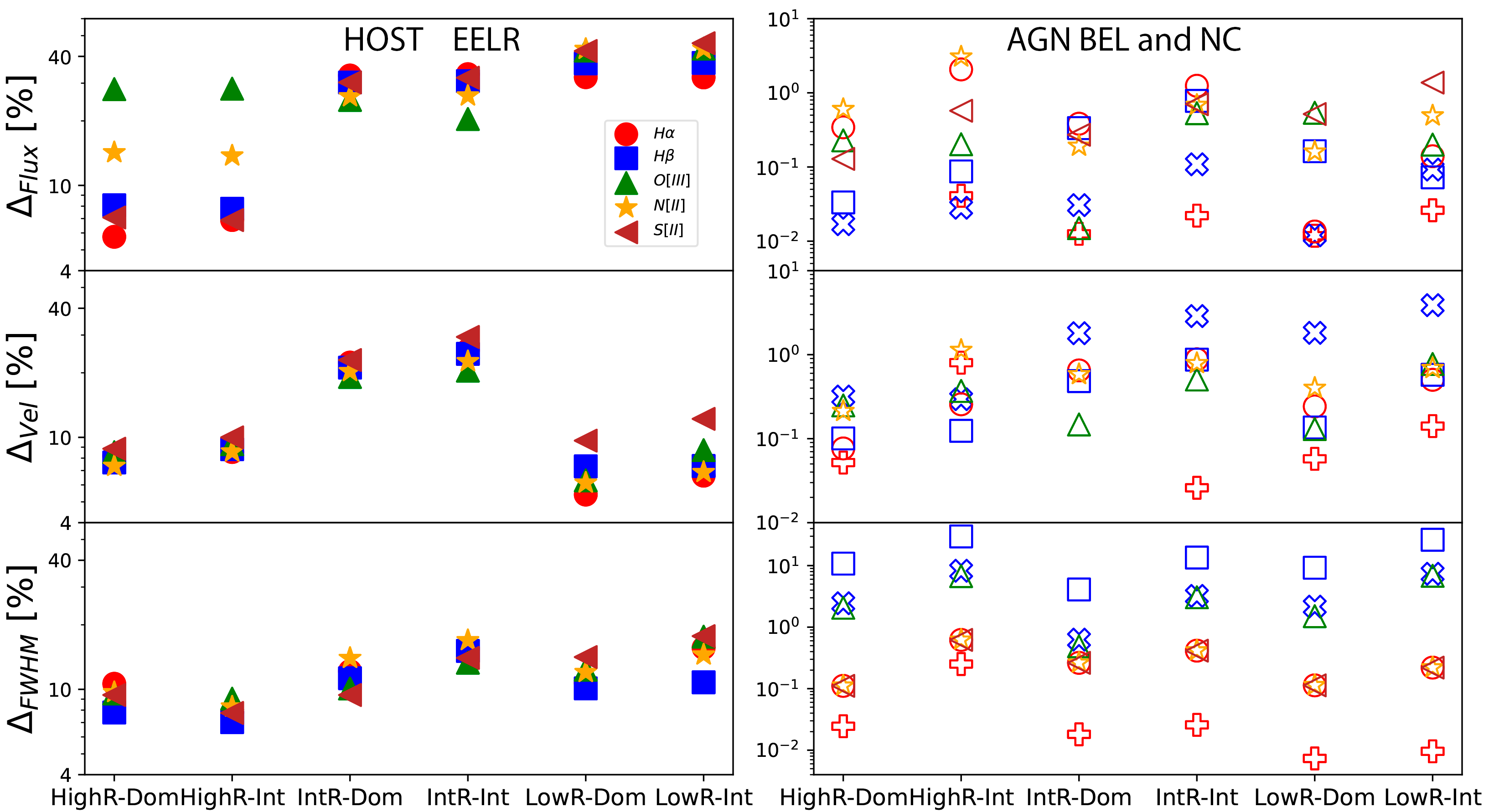}
\caption{Absolute values of the normalised residuals $\Delta val=|val_{in}-val_{out}|/val_{in}$, for the total integrated flux (upper panels), velocity shifts (middle panels) and the line FWHMs (lower panels). The left panels contain the flux-averaged $\Delta val$ within the central 3'' for the HG EELRs. The symbols represent the residuals for five emission lines: red filled circle for \ha, blue filled square for \hb, green filled up triangle for \oiiiopt, yellow filled star for \niill, and red filled left triangle for \siill. The right panels contain the $\Delta val$ for the AGN BEL and NCs lines: red circle for the narrow \ha, blue square for the narrow \hb, red cross for the broad \ha, blue X for the broad \hb, green up triangle for \oiiiopt, yellow star for \niill, and red left triangle for \siill.}
\label{fig:ELMockE}
\end{figure*}

\begin{table}
    \centering
    \begin{tabular}{lcccc}
         \hline\\
         Mock Type & Resolution & AGN/HG & \multicolumn{2}{c}{Average Normalised Residuals}\\
         & $R_e/spaxel$ & Flux Ratio & $> 5,000$\AA & $< 5,000$\AA\\ 
         \hline
         \hline
         HighR-Dom & 0.05 & 3.2 & 10$\%$ & 5$\%$  \\
         HighR-Int & 0.05 & 1.2 & 20$\%$ & 10$\%$ \\
         IntR-Dom  & 0.09 & 3.2 & 20$\%$ & 10$\%$ \\
         IntR-Int  & 0.09 & 1.2 & 25$\%$ & 15$\%$ \\
         LowR-Dom  & 0.13 & 3.2 & 15$\%$ & 5$\%$  \\
         LowR-Int  & 0.13 & 1.2 & 20$\%$ & 10$\%$ \\
         \hline\\
    \end{tabular}
    \caption{Average normalised residuals of the AGN-HG deblended spectra.}
    \label{tab:presicion_percentages}
\end{table}

With the six mock IFU spectra, we apply our AGN-HG decomposition algorithm and compare the deblended output HG and AGN spectra with the input HG and AGN spectra. With the input and output spectra, we calculate the flux residuals of the recovered decomposed spectra, defined as \begin{equation}\Delta Flux =(Flux_{out}-Flux_{in})/Flux_{tot},\end{equation} where $Flux_{out}$ is the recovered AGN or HG spectra, $Flux_{in}$ is the input AGN or HG spectra and $Flux_{tot}$ is the total contribution of the combined HG and AGN spectra. Therefore, $\Delta Flux$ is the residual between the input and output spectra normalised by the total observed flux of the spectra. In addition, by normalizing the spectra by $Flux_{tot}$, the absolute flux residuals of the recovered AGN and the recovered HG are the same. As we explain in Section \ref{sec:method}, the recovered HG spectrum is the residual of the original spectra minus the recovered AGN spectrum. Therefore, the sum of the recovered HG plus the recovered AGN spectra will always be the input spectrum. A summary of the average normalised residuals is given in percentages in Table \ref{tab:presicion_percentages}. In the next subsections, we describe our results about the AGN/HG decomposition of the spectra continuum, the deblendind of the emission lines of both the HG and the AGN spectra, and the recovering of the point-spread function as a function of the wavelength.

\subsubsection{Deblended spectra residual}

\begin{itemize}
    \item[a)] High-resolution AGN-dominated resolved spectra (HighR-Dom): We show in the upper panels of Figure \ref{fig:high_resolution} the results of the HighR-Dom mock decomposition. The Figure shows the total AGN+HG, the input HG, and the input AGN spectra (within 2.5 arcsecs of the IFU centre) as black, yellow, and blue solid lines each. The decomposed output HG and AGN spectra are each marked with green and red solid lines. Below the panel, we show as a solid black line the deblended normalised residual $\Delta Flux$. The solid grey line is the normalised error of the input spectrum defined as Error/Input. We recovered the AGN spectrum within a residual of $10\%$ of the total flux (AGN+HG) on the continuum above 5,000 $\angstrom$. On the other hand, below 5,000 $\angstrom$, the residual decrease to $5\%$ at 4,000 $\angstrom$. The normalised error of the spectrum is comparable with the residual of the recovered decomposed spectrum. 
    

    \item[b)] High-resolution intermediate AGN resolved spectra (HighR-Int): The bottom panels of Figure \ref{fig:high_resolution} show the results of the AGN-HG decomposition when the AGN flux is 1.2 times the HG flux. Compared to the HighR-Dom case, the method recovers the AGN spectrum with a residual of $20\%$ of the total flux greater than 5,000 $\angstrom$. Below that wavelength, the residual increases to $\approx 10\%$ at 4,000 $\angstrom$. The normalised error is comparable to the residual of the AGN spectrum. 
    

    \item[c)] Intermediate-resolution AGN-dominated resolved spectra (IntR-Dom): The upper panels of Figure \ref{fig:intermediate_resolution} show the flux of the deblended AGN spectrum when it is 3.2 times larger than the HG. In contrast to the HighR-Dom case, the AGN-deblended spectrum has a residual with a value of $20\%$ greater than 5000 $\angstrom$, and below that the residual tends to increase to $\approx 10\%$ at 4,000 $\angstrom$. In addition, the normalised error is below the values of the AGN-HG deblended residual. 
    

    \item[d)] Intermediate-resolution intermediate AGN resolved spectra (IntR-Int): The method returns an AGN deblended spectrum with a normalised residual of $25\%$ above 5,000 $\angstrom$. Bellow that wavelength, the residual increases to $\approx 15\%$ at 4,000 $\angstrom$ and $\approx 7\%$ at 3,500 $\angstrom$. In addition, the normalised error is between 3 to 5 times lower than the AGN spectrum residual. We show the results of the adopted parameters in the lower panels of Figure \ref{fig:intermediate_resolution}. 
    

    \item[e)] Low-resolution quasar-dominated resolved spectra (LowR-Dom): In this case, we explore an IFS observation with low spatial resolution with an AGN spectrum flux 3.2 times stronger than the HG flux. We show in the upper panels of Figure \ref{fig:low_resolution} the results of this case. Our deblended method is capable of deblending the AGN spectrum with a normalised residual of $15\%$ of the total flux greater than 5,000 $\angstrom$, and below that wavelength the residual increases to $5\%$ at 4,000 $\angstrom$. 
    

    \item[f)] Low-resolution intermediate quasar resolved spectra (LowR-Int): For this case, we present the results for a low spatial resolution IFS observation where the AGN spectrum flux is not dominant. We show these results in the lower panels of Figure \ref{fig:low_resolution}. The normalised residual of the deblended AGN spectrum is on the order of $30\%$ above 6,000 $\angstrom$. Between 5,000 and 6,000 $\angstrom$, the residual has a value of $\approx 20\%$. At 4,000 $\angstrom$, the residual reaches a value of $10\%$. 
    

\end{itemize}

\subsubsection{Emission lines deblended residuals}

To perform this comparison, we model and fit the HG EELR and the AGN BAL and NC from the input and output spectra for each mock case. For that aim, we use the {\sc pyPipe3D} analysis tool to fit the spatially resolved \ha, \hb, \niill, \siill\ and \oiiia\ HG emission lines, and the {\sc IRAF-specfit} to fit the AGN \ha, \hb, \niill, \siill\ and \oiiia\ NCs and the \hb\ and \ha\ BELs. With the emission line model, we obtain each line's total flux, velocity shift and the FWHM. Hence, we can compare how well our deblended method is able to recover those emission line properties. We define the parameter \begin{equation}\Delta val=\frac{|val_{in}-val_{out}|}{val_{in}},\end{equation} where $val_{in}$ represents the total flux, velocity shift or FWHM values estimated from the input HG or AGN spectra, and $val_{out}$ represents the value estimated from the deblended recovered spectra. Therefore, the parameter $\Delta val$ represents the absolute value of the normalised residual between the input and output values. Since estimating the input and output parameters uses the same fitting methodology and the same fitting constraints, $\Delta val$ indicates the absolute deviation of the recovered emission line properties from the deblended spectra compared with the true values. We describe the results for the HG EELR, AGN BEL, and NCs.

\begin{itemize}
    \item[a)] The HG EELR: We use the spatially resolved emission line values provided by {\sc pyPipe3D} to obtain the $\Delta val$ of the flux, velocity shifts and FWHM of \ha, \hb, \niill, \siill\ and \oiiia\ emission lines. In the left part of Figure~\ref{fig:ELMock} shows how the \ha\ $\Delta val$ is spatially resolved at different resolutions for the AGN dominated cases. Figure~\ref{fig:ELMockE} shows the flux average $\Delta val$ values integrated within the central 3'', which is the equivalent of the SDSS fiber diameter. The central flux values contain the larger residual of the HG EELR parameters. However, we see a small dependency on the spatial resolution of the IFS: for the HighR-Dom and HighR-Int cases, the flux $\Delta val$ ($\Delta flux$) for \ha, \hb\ and \siill\ are between the $6\%$ to $9\%$ of the input flux, with the  \niill\ around $13\%$ and  \oiiia\ around $30\%$ of the input fluxes. However, for the IntR-Dom and IntR-Int cases, all the lines have a $\Delta flux$ around $30\%$, and for LowR-Dom and LowR-Int, we obtain a $\Delta flux$ around $40\%$.
    
    For the velocity shift ($\Delta vel$), we cannot find a dependency with the IFS spatial resolution, but we find a much lower $\Delta val$ than the flux case. For the HighR-Dom and HigR-Int cases, we find a $\Delta vel$ around $8\%$. In contrast, the IntR-Dom and IntR-Int recuperate a $\Delta vel$ around $30\%$, and the  LowR-Dom and LowR-Int obtain a value around $8\%$. Finally, we also find a weak dependency with the IFS spatial resolution for the FWHM $\Delta val$ ($\Delta FWHM$). For the HighR-Dom and HigR-Int cases, the $\Delta FWHM$ values are between $7\%$ to $9\%$, for the IntR-Dom, IntR-Int, LowR-Dom and LowR-Int cases the cases the value of $\Delta FWHM$ are around $11\%$ to $15\%$.

    As shown in Figure~\ref{fig:ELMock}, the central region of the galaxy is the most affected by the AGN emission. However, we note that the larger the spatial resolution, we obtain a lower deviation of the true values is extended. This result is expected since the larger spatial resolution implies a small relative size of the PSF FWHM compared to the IFS spatial sampling. It is also important to note the asymmetries of the residuals on the $\Delta flux$ for all the cases. This result shows the method's difficulty in dealing with asymmetries when modelling the HG SB. We plan to improve our method by adding more sophisticated models that include bar and disk structures. The maps also show that the velocity shift and the line FWHM are well recuperated within a more confined region. In addition, as Figure~\ref{fig:ELMockE} already shows, the Intermediate resolution case is the one that has the highest $\Delta vel$ of the other two cases. The high residual value of $\Delta vel$ can also be related to the HG SB.
    
    \item[b)] The AGN BEL and NC fitting: We use the {\sc IRAF-specfit} routine \citep{kriss94} to fit the input and output AGN spectra. The spectral fitting considers the peculiarities in the shape of the input AGN line profiles. We consider the emitting region that comprises the \hb\ region, including the \oiiiopt\ doublet and the emission of the pseudo continuum of \feii. For the AGN input spectra (see Section \ref{sec:Mock IFU Spectra}), it is of particular importance the \feii\ emission, which can affect the estimation of the emission from the red wing of \hb\ \citep[e.g.][]{Kovavcevic+2010}. A wide window with intense \feii\ multiplet emission is required to constrain it adequately. We chose the window 4450-5400 \AA\ and the \feii\ template used in \citet{negrete2018}, which adequately reproduces the \feii\ emission around \hb. We tested two fits considering a Gaussian and a Lorentzian profile 
    for the broad \hb\ component to choose the one that met the requirements seen in the diagnostic diagram for quasars in the optical \citep[e.g.][figures 1 and 3 to 8]{negrete2018}. For a FWHM < 4000 km s$^{-1}$ for the BC of \hb, a Lorentzian profile should be used. The choice of this profile is also supported by the moderate \feii\ emission and returns the best fit with a lower $\chi^2$ than using a Gaussian profile. For the remaining components, the \hb\ narrow component and the \oiiiopt\ doublet, we use Gaussian profiles with the same shift and FWHM. As reported in \citet{negrete2018}, we considered a second component for \oiiiopt\ with a free shift and FWHM on the basis that it is a high ionization line and is susceptible to showing blue asymmetries.
    
    In the \ha\ region, we perform a spectral fitting in the 6200-6850 \AA\ range that comprises the intense broad \ha\ fitted with the same line profile as \hb, and the narrow components \niill\ and \siill\ fitted again with Gaussian profiles, sharing the same shift and FWHM. For both \ha\ and \hb\ fits, we consider the same (within the errors) continuum fitted with a single power law. Once all the components were considered, we made the fitting with {\sc IRAF-specfit} via $\chi^2$ minimization using a Marquardt algorithm \citep{kriss94}.
    
    We show the results on the right part of Figure~\ref{fig:ELMockE}. Because of the intense emission of the AGN BELs, the $\Delta flux$ of the \ha\ and \hb\ broad components have the lowest residuals as expected with values around $0.01\%$ to $0.1\%$. For the case of the $\Delta vel$ of the broad components, we find a wide range of residuals, however they are almost negligible in comparison with the HG EELR counterparts, with values around $0.01\%$ to $4\%$. For the $\Delta FWHM$ of the broad components, we find a value range between $0.01\%$ to $9\%$, with \ha\ being more consistent with a residual value around $0.01\%$ and \hb\ with a value around $9\%$. In any case, we cannot find any dependency on the IFS resolution or the dominance of the AGN-HG emission.
    
    For the case of the NCs, we find that the $\Delta flux$ residuals have values around $0.3\%$. Specifically, for the HighR-Dom and HighR-Int cases, \ha NC has the largest residuals around $1\%$ and \hb NC the lowest residuals around $0.08\%$ with \oiiiopt, \niill\ and \siill\ scattered between those values. For the IntR-Dom and IntR-Int cases, all lines have a $\Delta flux$ residual around $0.3\%$; and for LowR-Dom and LowR-Int, the residual values are scattered between $0.01\%$ and $1\%$ with \ha\ and \hb\ NCs having the lowest residuals. Similarly, we cannot find any dependency on the IFS resolution or the dominance of the AGN-HG emission. However, for the case of $\Delta vel$, we have a weak dependency on the AGN intensity. We measure a small residual value for the AGN-dominated cases compared to the AGN intermediate case. We cannot find any clear dependency with the residuals for the spatial resolution. Overall, $\Delta vel$ is around $0.3\%$ for all the lines with a range from $0.1\%$ to $1\%$. Finally, for the $\Delta FWHM$ residual, we have a more consistent value among all the mock cases without any clear dependency. The \ha NC, \niill, and \siill\ emission lines have the lowest residuals around $0.2\%$, with \hb NC and \oiiiopt being the ones with the largest residual around $1\%$.
\end{itemize}

\begin{figure}
\includegraphics[width=\columnwidth]{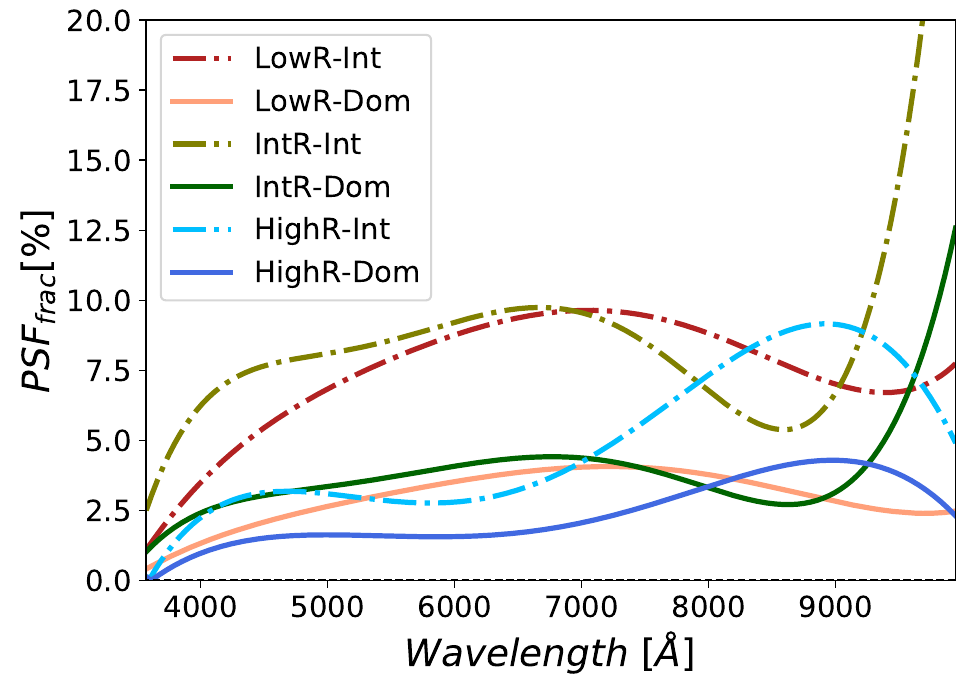}
\caption{Debledended FWHM PSF size normalised residual as a function of wavelength. The solid lines represent the Quasar-dominated cases (Dom), and the segmented lines represent the Intermediate quasar-dominated cases (Int).}
\label{fig:psf_deltas}
\end{figure}

\begin{figure}
\includegraphics[width=\columnwidth]{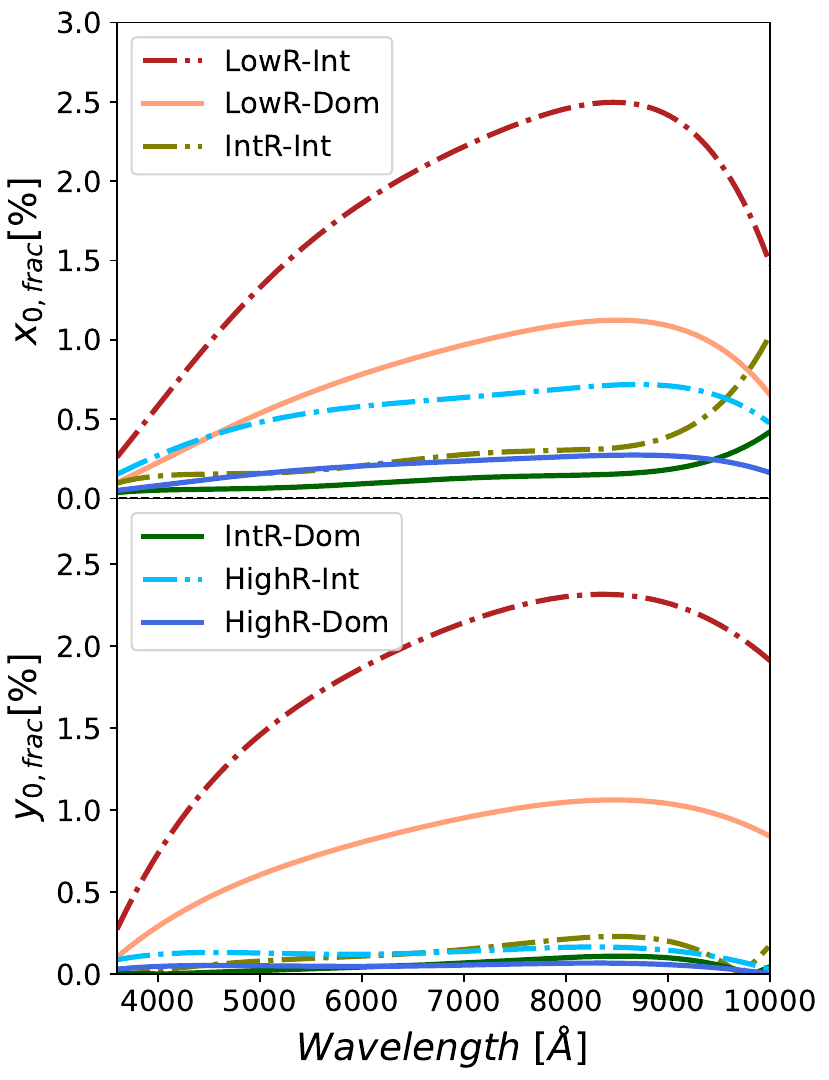}
\caption{Debledended PSF $x0$ and $y0$ normalised residuals as a function of wavelength. We use the same colour code as in Figure \ref{fig:psf_deltas}.}
\label{fig:xy_deltas}
\end{figure}

\begin{figure*}
\includegraphics[width=2\columnwidth]{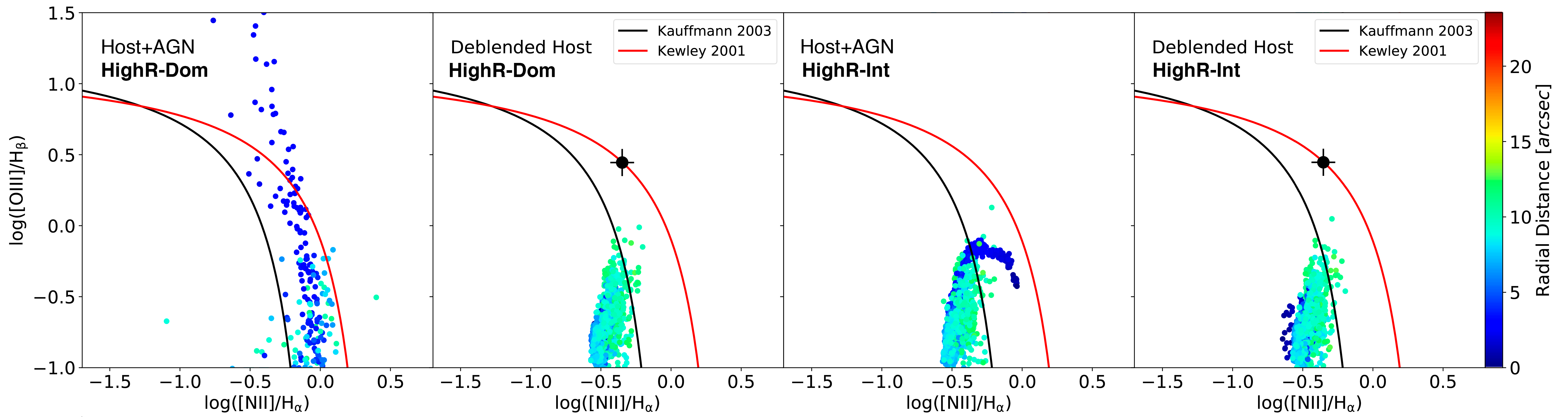}
\includegraphics[width=2\columnwidth]{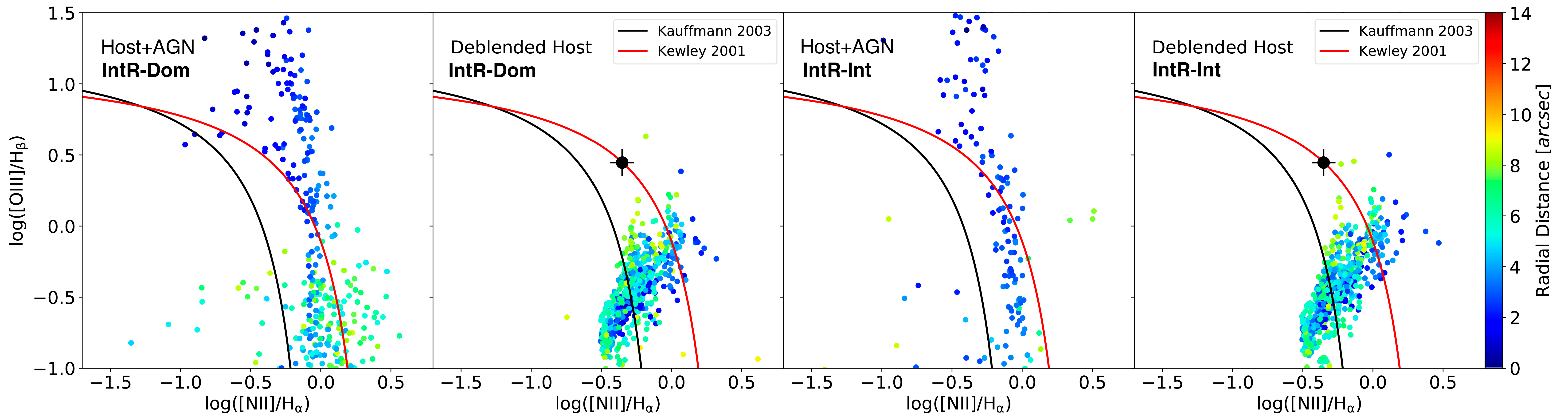}
\includegraphics[width=2\columnwidth]{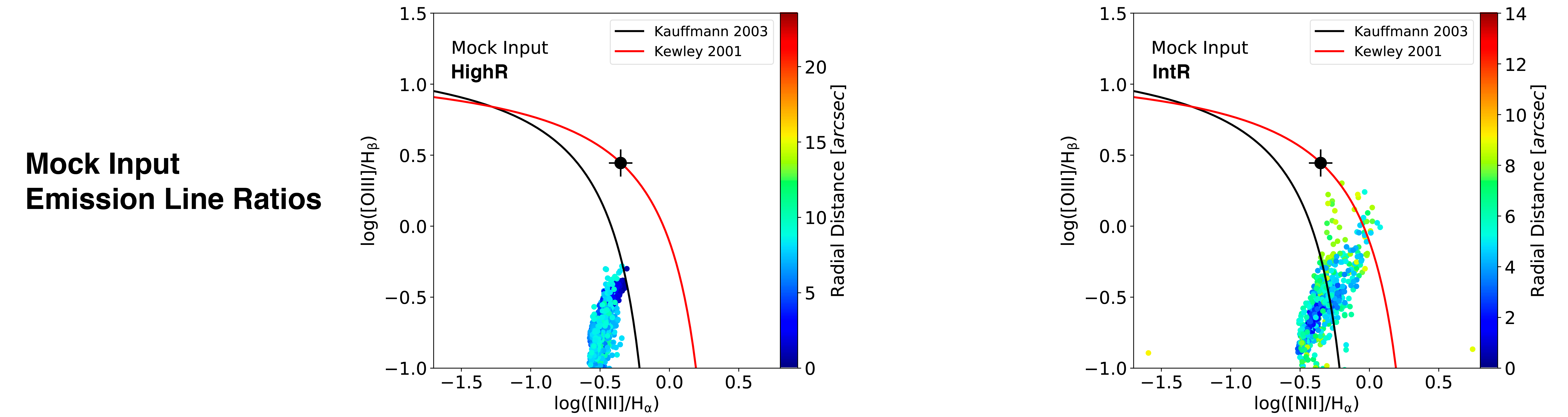}
\caption{The Baldwin, Philips \& Terlevich NII diagram for the HighR (top), IntR (middle) Manga mock galaxies and its host galaxy input values (bottom). Top and middle panels from left to right: The spatially resolved BPT diagrams for the mock AGN+HG and the deblended host of the HighR-Dom,  HighR-Int,  IntR-Dom, and IntR-Int cases. Bottom panels from left to right: The spatially resolved BPT diagrams of the input host galaxies of the HighR case (left) and the IntR case (right). The colour bar represents the radial distance from the IFU central position. The solid red line represents the \citet{Kewley+01} demarcation line. The solid black line represents the \citet{Kauffmann+2003May} demarcation line. We plot only the spaxels with an SNR greater than 5. The solid black point represents the BPT position of the input quasar composite spectra.}
\label{fig:BPTMock}
\end{figure*}

\begin{figure*}
\includegraphics[width=2\columnwidth]{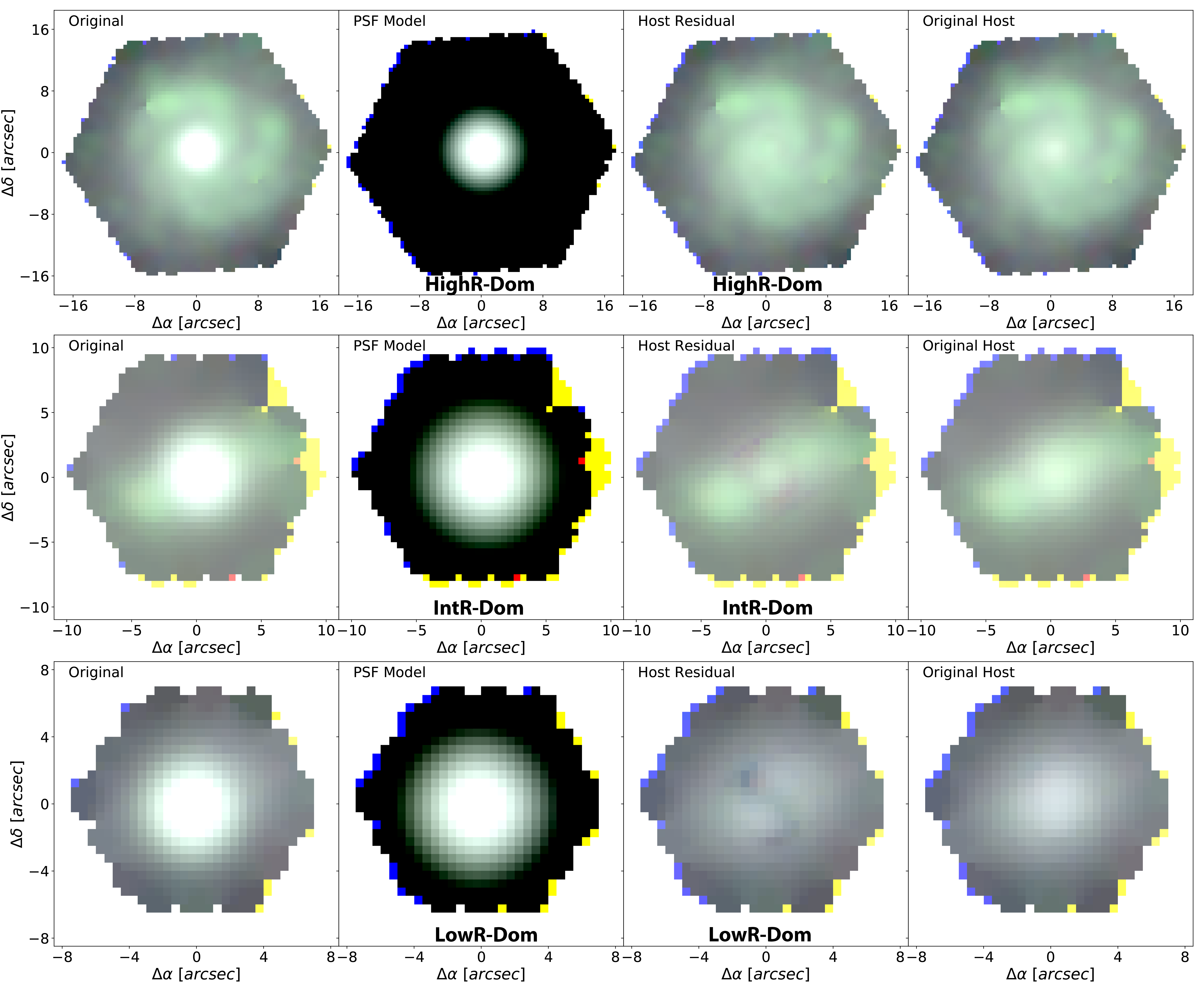}
\caption{Composed RGB image with B-\oiiia~, G-\ha~ and R-\niib maps for the MaNGA mock galaxies for the AGN dominated cases: HighR-Dom (top), IntR-Dom (middle), and LowR-Dom (bottom). The left panels show the non-deblended maps, the central left panels show the deblended PSF maps, the central right panels show the deblended host map, and the right panels show the input HG map used in the mocks.}
\label{fig:map_example_rgb}
\end{figure*}

\begin{figure*}
\includegraphics[width=2\columnwidth]{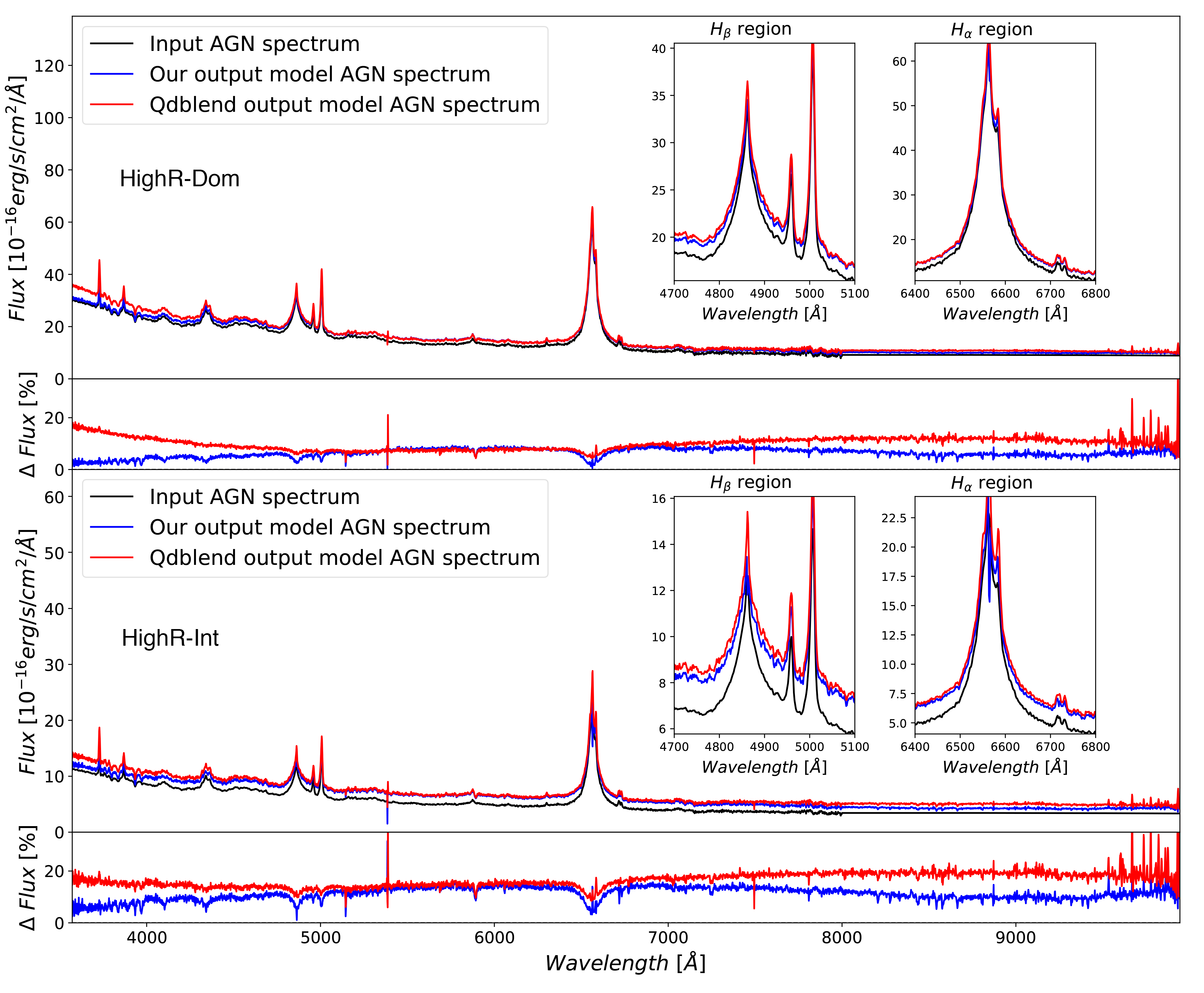}
\caption{{\sc QDeblend3D} comparison of our deblended AGN spectra for the HighR-Dom case (upper panels) and HighR-Int case (lower panels). The solid blue line represents the deblended quasar spectra obtained with our methodology, the solid red line represents the AGN spectra obtained with {\sc QDeblend3D}, and the solid black line represents the input AGN spectra of the mocks. The two inset plots show a zoom-in of the $H_{\beta}$ and $H_{\alpha}$ regions. The small panels below each principal panel show the normalised residuals: The {\sc QDeblend3D} (red) and our deblendig spectra (blue) minus the input AGN spectra over the input AGN spectra. The spectra are the total flux within 2.5 arcsecs of the centre of the IFU-FoV.}
\label{fig:qdblendMock}
\end{figure*}

\subsubsection{Deblended PSF precision}

The residuals in the deblending decreases as the AGN flux is more intense with respect to the HG flux. In all the cases, the residuals always decreases at lower wavelengths, where the nonthermal power-law continuum of the AGN increases and blurs the flux of the HG. This effect is also seen in the BELs, where the total flux of the AGN is larger than the HG. On the other hand, the IFS spatial resolution has an impact on the deblending of the HG EELRs: the larger the spatial resolution, the better the deblending of the HG emission lines. In this work, we define the spatial resolution as the relative size of the FWHM of the PSF in comparison with the spaxel size: in other words, the spatial resolution is the minimum size on an IFS from which it is possible to resolve two different structures, that differs from the spatial sampling (the spaxel size). Therefore, our method is sensitive to the total resolution within the IFS FoV. It is important to note that when the spatial resolution decreases, the FoV of the IFS in proportion to the FWHM size of the PSF also decreases. This behaviour is due to the fact that (at least for homogeneous IFS surveys) the FWHM of the PSF remains stable for all the IFS observations: the instrument is always at the same observatory. Therefore, there is an error in the estimation of the PSF Moffat wings when the resolution of the IFS observation changes. To explore this issue, we measure how the FWHM size of the recuperated PSF changes relative to the input values for all mock cases. In this point, we use the definition of the Moffat FWHM PSF profile of \citet{Trujillo+2001}, that is $FWHM_{PSF}=2\alpha\sqrt{2^{1/\beta}-1}$. We can define the normalised FWHM PSF residual size as: \begin{equation}PSF_{frac}=\frac{FWHM_{PSF,in}-FWHM_{PSF,out}}{FWHM_{PSF,in}},\end{equation} where $FWHM_{PSF,in}$ is the input FWHM size of the PSF used to create the mock AGN spectra as described in Section~\ref{sec:Mock IFU Spectra}, and $FWHM_{PSF,out}$ is the output FWHM size of the PSF of the deblended AGN spectra.

We show the values of $PSF_{frac}$ along the wavelength range for all six mock cases in Figure \ref{fig:psf_deltas}. For all the cases where the AGN flux spectrum domains over the HG spectrum (HighR-Dom, IntR-Dom, and LowR-Dom ), the residuals of the recovered FWHM size of the PSF is within $5\%$ below 8,500 $\angstrom$. Above that wavelength, the errors associated with the imperfect subtraction of the skylines (on the original IFS observations) start to degrade the PSF modelling. As a consequence, the final FWHM size of the PSF starts to depart from the input values. On the other hand, for the IntR-Int, and LowR-Int cases, the residuals of the deblended PSF is within $15\%$. The results indicate that the relative flux between the non-resolved and the resolved emission (AGN/HG) has a larger impact on accurately modelling the PSF profile. This is expected because the larger the contribution of the AGN is, the easier it is for the method to model the PSF. However, only for the HighR-Int case, the residuals are also within $5\%$, indicating that when a sufficient spatial resolution is available, it is also possible for the method to model the PSF profile accurately even if the AGN emission is not too bright.

So far, we have described the method performance to recuperate the input FWHM of the non-resolved PSF SB. However, the FWHM only traces the recovering performance of the input $\alpha$ and $\beta$ parameters \citep{Trujillo+2001}. To explore the recovering performance of the PSF centroid ($x_0$ and $y_0$), we define the next smoothed normalised residuals: \begin{equation}x_{0,frac}=\frac{x_{0,in}-x_{0,out}}{x_{0,in}}, \\ y_{0,frac}=\frac{y_{0,in}-y_{0,out}}{y_{0,in}},\end{equation} where $x_{0,out}$ and $y_{0,out}$ are the recovered and smoothed spectral values of $x_{0}$ and $y_{0}$. In Figure~\ref{fig:xy_deltas}, we show the results for all our mock cases. The normalised residuals show that $x_0$ and $y_0$ are the best-recuperated parameters of the PSF profile, whatever the mock case. For the HighR-Dom, IntR-Dom and IntR-Int cases, the normalised residuals are below $0.5\%$, especially for $y_0$, where the normalised residuals are below $0.3\%$. For the HighR-Int case, the maximum residual values are below $0.7\%$ for $x_0$ and below $0.3\%$ below $y_0$. Finally, for the LowR-Dom and LowR-Int, the maximum residual values reach $1\%$ and $2.5\%$  for each case for the $x_0$ and $y_0$ parameters. Therefore, while the centroid of the PSF is the best-recuperated parameter, there is a dependence on the spatial resolution: a large spatial resolution returns a lower normalised residual.

\subsection{The BPT Diagram on the deblended HG spectra}\label{Mock_BPT}

We now explore the results of the deblending of the HG EELRs returned by our method. For this case, we use the Baldwin, Philips \& Terlevich NII diagram \citep{Baldwin+81}. The BPT diagram is one of the most known narrow-line diagnostic diagrams in the literature \citep[e.g.,][]{Veilleux+87,Kewley+01,Kauffmann+2003May}. This diagram is used to disentangle the ionisation properties of the nebular gas that can be associated with AGN emission, star formation, and aged stars. Therefore, the BPT diagram is ideal for testing the AGN/HG deblending of the EELRs. In this exploration, due to the size of the FoV of the low-resolution cases (LowR-Dom, LowR-Int), we use only the deblending results for the high- and intermediate-resolution cases. In Figure \ref{fig:BPTMock} we show the BPT diagrams before and after the deblending process. For this analysis, we use {\sc pyPipe3D} to analyse the nebular emission lines for the original and HG+AGN  data cubes of our three galaxies. It is important to note that {\sc pyPipe3D} is not designed to analyse broad emission line spectra; however, it performs a well-suitable continuum subtraction to analyse the EELRs. Therefore, we can use the nebular line analysis of {\sc pyPipe3D} to generate the BTP diagram of the non-deblended spectra only for comparison. In addition, to construct the spatially resolved BPT diagram, we use only the spaxels that have an SNR greater than 5. 

For the case of the non-deblended HighR-Dom, the resolved BPT shows intense contamination due to the AGN emission within the central 5 arcsec. The central spaxels show an emission narrow line ratios originating from an AGN ionisation source: the line ratios lie above the \citet{Kewley+01} demarcation line. In addition, the middle spaxels (beyond 5 arcsecs) lie below the Kewley demarcation line but above the \citet{Kauffmann+2003May} line. The Kauffmann and Kewley lines define the composite region, where it is not completely clear if the main ionisation source comes from an AGN or a star formation region. On the other hand, the spatially resolved BPT diagram suffers a radical change when the AGN emission is deblended. After removing the nonresolved AGN emission, almost all the line ratios of the HG EELRs lie below the Kauffmann demarcation line. This result shows that the only ionisation sources are from star formation regions, as expected from the input mocks. 

For the non-deblended HighR-Int case, we obtain that most of the central spaxels lie in the BPT composite region. However, after applying the HG/AGN deblending algorithm, most of the spaxels are in the star-forming region. Once we compare with the input values of the mock host galaxies (before adding the AGN spectra), the method can recover the true host emission line ratios, with only a small number of spaxels slightly crossing the Kauffman line.  

For the non-deblended IntR-Dom and the IntR-Int cases, we obtain a very similar BPT spaxel distribution as the HighR-Dom case, with most of the central spaxels lying above the Kewley line or within the composite region of the BPT. On the contrary, the BPT diagram estimated from the deblended HG spectra shows a completely different behaviour. Most of the central spaxels lie in a very well-defined region located between the BPT composite region and the star-forming region. Once we compare the line ratios of the input HG, we show that the BPT spaxel positions of our deblended HG spectra are the same as the input HG emission line ratios.

To visualise the spatially resolved AGN deblending across the BPT, we present in Figure \ref{fig:map_example_rgb} the Red-\niib, Green-\ha, and Blue-\oiiia~ in an RGB image of the input, the PSF model, and the residual map for the AGN/HG dominant cases. We also show the input HG map. For the HighR-Dom case, the method can spatially disentangle the nonresolved AGN emission from the resolved HG EELR. In its residual, it is clearly shown that most of the HG EELR is located on the disk of the galaxy, which agrees to be associated with SF regions. In a similar way, we find a similar result for the case of the IntR-Dom case, with the HG EELR clearly decoupled from the nonresolved AGN emission. Also, Figure \ref{fig:map_example_rgb} shows that the HG \ha~ emission is stronger than the \niib~ emission, as we show in Figure \ref{fig:BPTMock}. Finally, for the LowR-Dom case, we show that most of the \ha~emission comes from the AGN emission, as we also show in Fig. \ref{fig:low_resolution}. 

\subsection{Comparison with QDeblend3D}\label{qdb1}

In this section, we compare our deblending methodology with the one presented by the {\sc QDeblend3D} quasar deblending code presented by \citet{Husemann+13} and \citet{Husemann+22}. We decided not to compare our method with {\sc galfit3d} or {\sc q3dfit} because those methods require a prior characterization of the PSF profile, making a crucial difference with our method. On the other hand, {\sc QDeblend3D} does not perform a prior PSF modelling,  but it requires initial information on the relative flux of the AGN/HG. Therefore, the closest comparison of our method is with {\sc QDeblend3D}. We use only the HighR mocks because they have the best decomposition results: any deviation of the deblended spectra from the input spectra will be caused only by the deblending procedure and not by any limitation due to the IFS observation and PSF/FOV sampling.

Figure \ref{fig:qdblendMock} shows the deblending comparisons for the HighR-Dom and HighR-Int cases. For both cases, {\sc QDeblend3D} recovers with great detail the AGN continuum above 5,000 \AA\ and maintains a constant precision of $\approx$ 10$\%$ of the flux for the HighR-Dom case and $\approx$ 20$\%$ for the HighR-Int case. This precision is comparable with the recovered precision of our method between 5,000 \AA\ and 7,000 \AA. However, wavelengths below 5,000 \AA\ {\sc QDeblend3D} tend to return a bluer spectrum, increasing (or making it constant) the residual precision to 20$\%$. For wavelengths above 7,000 \AA\, the {\sc QDeblend3D} precision tends to be constant, while our methodology varies depending on the spectra error and the sky residuals. In addition, our method tends to return lower residuals for the same wavelength range. It is important to say that in the loci of the strong emission lines (\ha\ and \hb), while the residuals of the BEL are smaller in our method, the residuals of the NC are smaller and best recuperated by {\sc QDeblend3D}.

\subsection{Single Stellar Populations}

\begin{figure}
\includegraphics[width=\columnwidth]{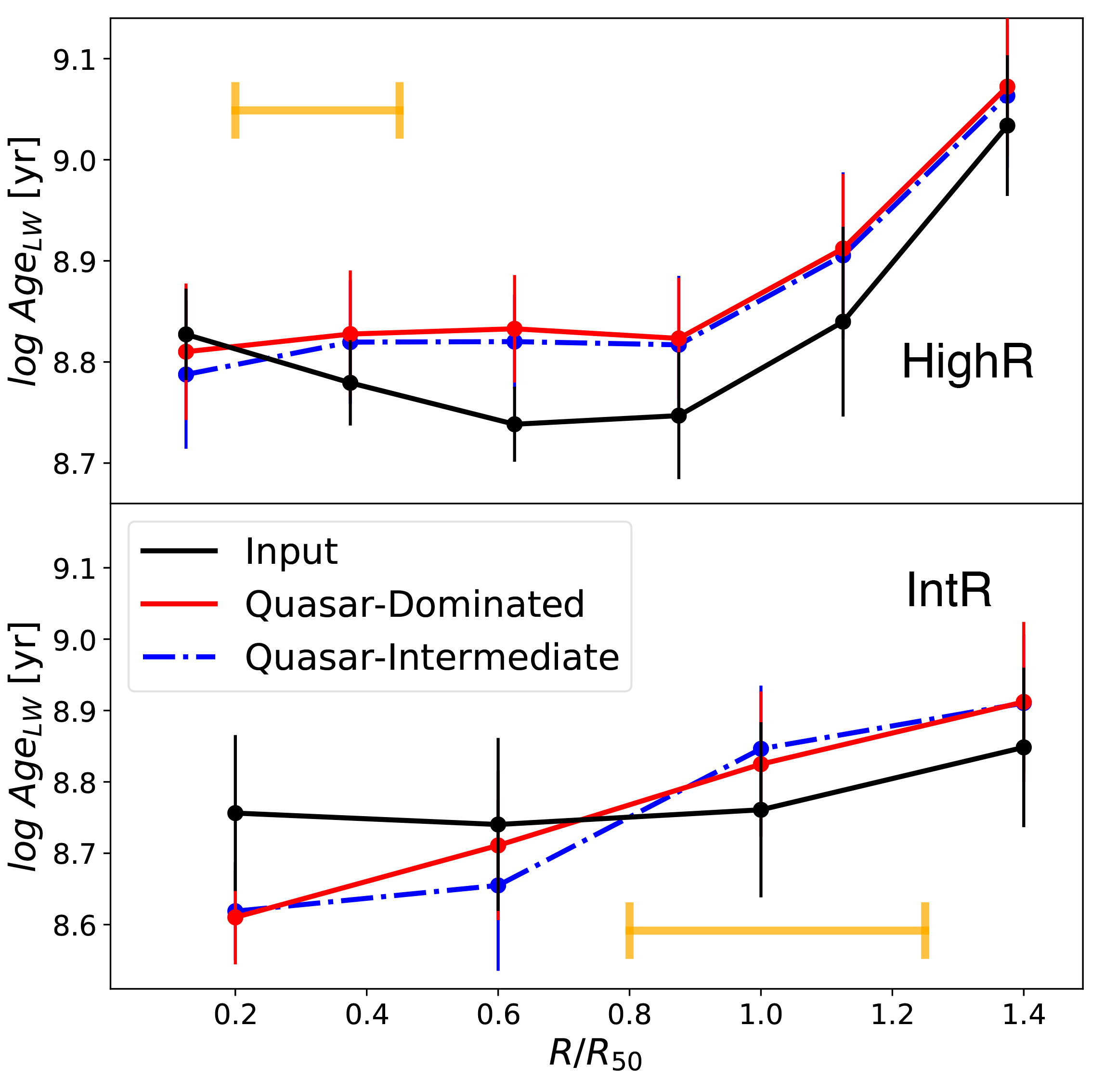}
\caption{Radial profiles of the Age$_{LW}$ for the high-resolution (upper panel) and intermediate-resolution (lower panel) cases. The black solid lines represent the radial profiles of the input MaNGA galaxies without the simulated AGN spectra. The red solid lines represent the radial profiles of the HG-deblended spectra for the quasar-dominated case. The segmented blue lines represent the radial profiles of the HG-deblended spectra for the quasar intermediate case. The yellow solid line represents the FWHM PSF size.}
\label{fig:ages}
\end{figure}

In this section, we explore the recovery of the SSP properties from the deblended HG spectra during phase two of our method. In particular, we explore the light-weighted (LW) SSP Ages ($Age_{LW}$) radial profiles. We will present a detailed analysis of the SSP properties in a forthcoming paper. For this test, we use the mock IFS cubes considered above. We select the $Age_{LW}$ because it is the most sensitive SSP property with respect to the slope of the spectra. Hence, any bias with respect to the HG deblending will directly impact the estimation of $Age_{LW}$. The definition of $Age_{LW}$ is described in detail in Equation 7 of \citet{Ibarra-Medel+22b}. However, we briefly define $Age_{LW}$ as the average of the SSP Ages weighted by the SSP spectral flux within a certain spectral window. In this case, we define the spectral window within 4,500 to 5,500 $\angstrom$.

To obtain the original LW Ages, we apply {\sc pyPipe3D} to the input MaNGA cubes before the construction of the mock data cubes. This step returns an SSP synthesis without the existence of a central bright AGN emission. We use the SSP analysis of the respective mock HG deblended spectra. Then, we obtain a direct comparison of the input SSPs (original data cube) with the deblended version of the same galaxy (deblended HG). For this test, we use only the HighR and IntR cases. The low spatial resolution of the LowR case cannot be used to obtain a good representation of the real Age profiles due to the size of the PSF, as we explored in \citet{Ibarra-Medel:2019aa}. 

In Figure \ref{fig:ages}, we show the results of these two cases.  For the HighR case, we find that the central Age values match those values of the input galaxy. For values larger than 0.6 $R_{50}$, the estimated Ages are 0.1 dex older than the original values. However, this difference is between the Age error estimations. For the IntR case, the SSP Ages of the central values return to be $\approx$ 0.2 dex younger than the Ages of the input galaxy, and then, at larger radii, the values tend to be older by $\approx$ 0.1 dex. Again, these differences are within the Age errors. For all of the cases, we find that the intensity of the AGN does not affect the SSP analysis of the deblended HG spectra.

Finally, it is important to note that a bias may exist in the SSP decomposition. As shown in Figure~\ref{fig:qdblendMock}, the residuals show that {\sc QDeblend3D} returns a more flattened spectral residual slope in contrast with our method. In addition, 
\citet{Sarmiento+2023} 
showed that the extinction parameter is mainly affected by the spectral slope. Therefore, the recovered dust extinction from the SSP decomposition can be biased to lower values for our deblended HG spectra. However, the age values are less affected by the spectral slope due to the SSP modelling, which also considers the absorption features, remaining within 0.2 dex of the true values, as shown in \citet{Sarmiento+2023}.

\subsection{Impact of the Mock Morphologies}

The selection of the input host galaxy spectra 1-211017 (HighR), 1-38348 (IntR) and 1-458316 (LowR) were chosen for their star formation activity and the condition they do not have any evident nuclear activity. However, they were not selected 
based on their morphological types. Their morphologies are Sb, SBb and SAB0a for 1-211017, 1-38348, and 1-458316, respectively. The morphologies were taken from the MaNGA Visual Morphologies from SDSS and DESI images Value Added Catalog\footnote{\url{https://www.sdss4.org/dr17/data_access/value-added-catalogs/?vac_id=manga-visual-morphologies-from-sdss-and-desi-images}} (Vazquez-Mata et al., in prep). 1-211017 and  1-38348 are spiral galaxies, with 1-38348 containing a bar structure. In addition, 1-458316 is a spheroid-like galaxy with a faint bar structure; see Figure~\ref{fig:MANGA}. The morphologies of the input HG for the HighR and LowR galaxies present structures that can not be well-modelled by a simple S\'ersic profile. For the LowR case, a simple S\'ersic profile is more suitable due to its morphology.

On the other hand, the main objective of our decoupling method is to disentangle the non-resolved (AGN) from the resolved (HG) flux by modelling the SB. To model the non-resolved flux, we require a background value to fit the PSF parameters. In addition, the actual spatial resolution and the field of view (FOV) of an IFS observation do not reach the values required that different morphologies could impact the performance of the method. Therefore, assuming a S\'ersic profile to model the resolved SB is a good option. The results confirm this: the best decoupling results are for the HighR, a typical Sb galaxy with complex spiral arm structures that cannot be well modelled with a simple S\'ersic profile. We obtain the largest residuals for the LowR case when a simple S\'ersic profile is more suitable due to its morphology. In addition, we will improve the code in a future version to model more complex structures of the host SB to deal with the upcoming high-spatial resolution IFS observations.

\subsection{Extra Interpolation Values}

Finally, we check the case when the $R_e$ and $I_0$ are set to be interpolated during the full spectral modelling phase of our method. We obtain that the fitting spectral response of $R_e$ presents small fluctuations per wavelength slide. Those fluctuations are originated by very small changes of the SB fitting due to subtle noise level changes on the data and the stochastic behavior of an MCMC fitting. However, those fluctuations per spectral pixel are strongly reduced with the interpolation. On the other hand, $I_o$ principally traces the host galaxy spectra of all the emissions and absorptions not associated with the AGN emission, and it is more affected by the original noise from the original spectra.  However, as discussed in the previous section, the decoupling of the non-resolved spectra is not affected since the method only uses the HG model as a background value, and the final recovered HG spectrum is simply the residual of the original spectra minus the deblended PSF spectra.


\begin{figure*}
\includegraphics[width=2\columnwidth]{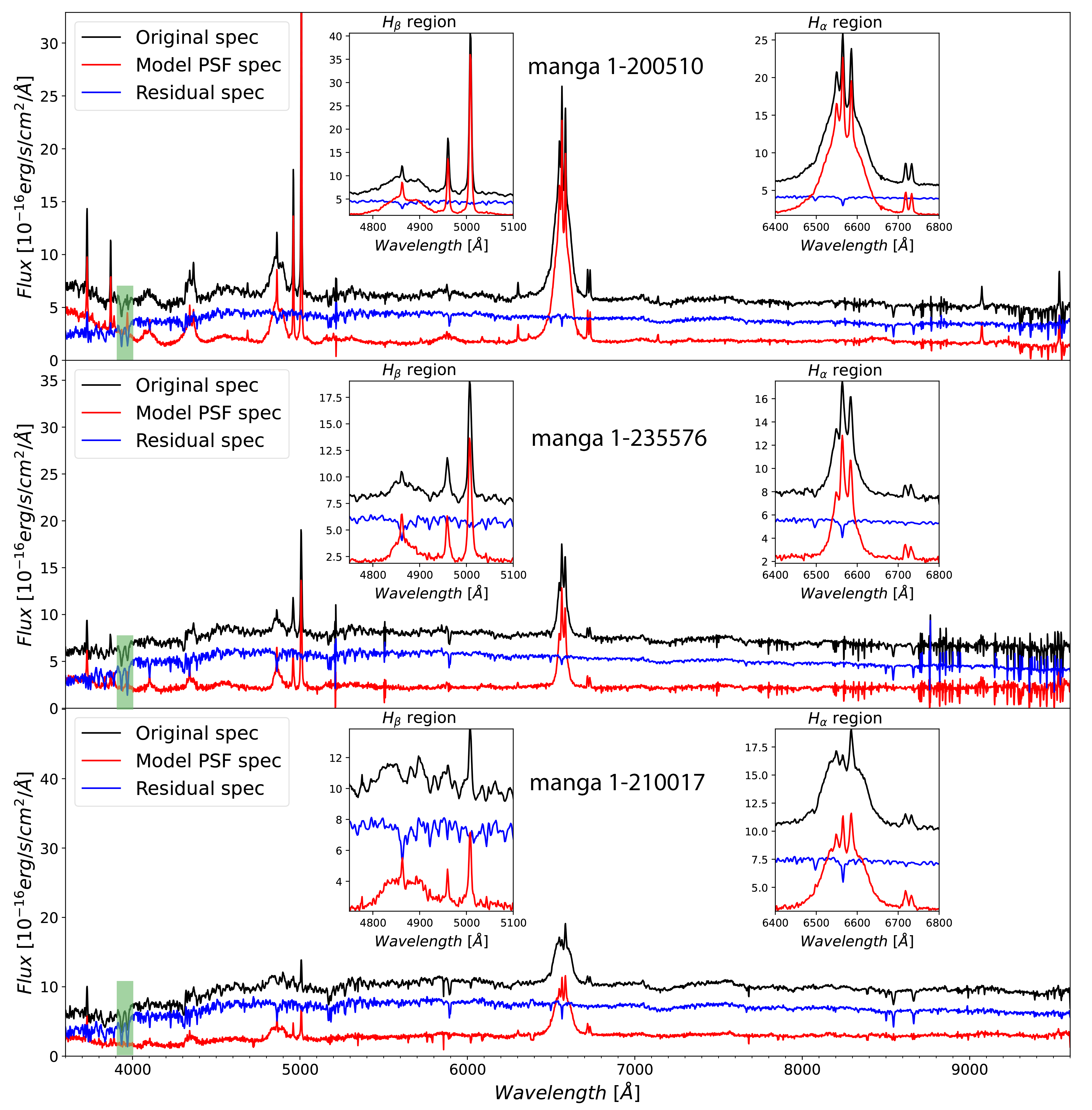}
\caption{Deblended spectra for the MaNGA galaxies 1-200510 (top), 1-235576 (middle), and 1-210017 (bottom). For all panels, the solid black line represents the input spectra (HG+AGN), the solid red line represents the deblended AGN spectra, and the blue solid line represents the deblended HG spectra. The two inset plots show a zoom-in of the $H_{\beta}$ and $H_{\alpha}$ deblended quasar spectra. The spectrum is the total spectrum within 2.5 arcsec of the centre of the IFU-FoV. The green rectangles mark the wavelength location of the  HG stellar {C\sc{II}} HK absorption lines.} 
\label{fig:spectral_debl}
\end{figure*}

\begin{figure*}
\includegraphics[width=2\columnwidth]{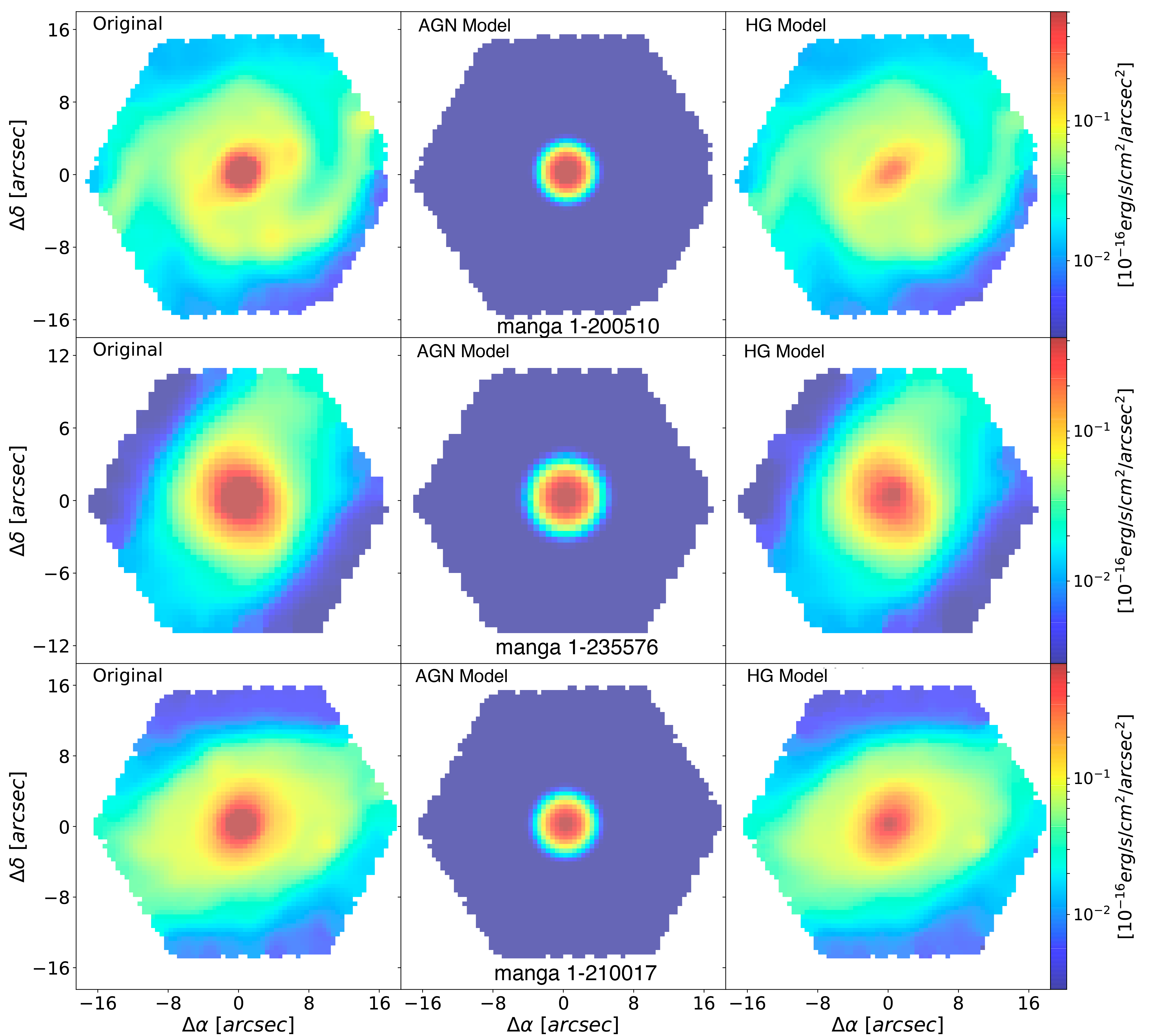}
\caption{\ha~surface brightness maps for the MaNGA galaxies 1-200510 (top), 1-235576 (middle), and 1-210017 (bottom). The left panels show the non-deblended maps, the central panels show the deblended PSF maps, and the right panels show the deblended host map.}
\label{fig:map_example}
\end{figure*}

\begin{figure*}
\includegraphics[width=2\columnwidth]{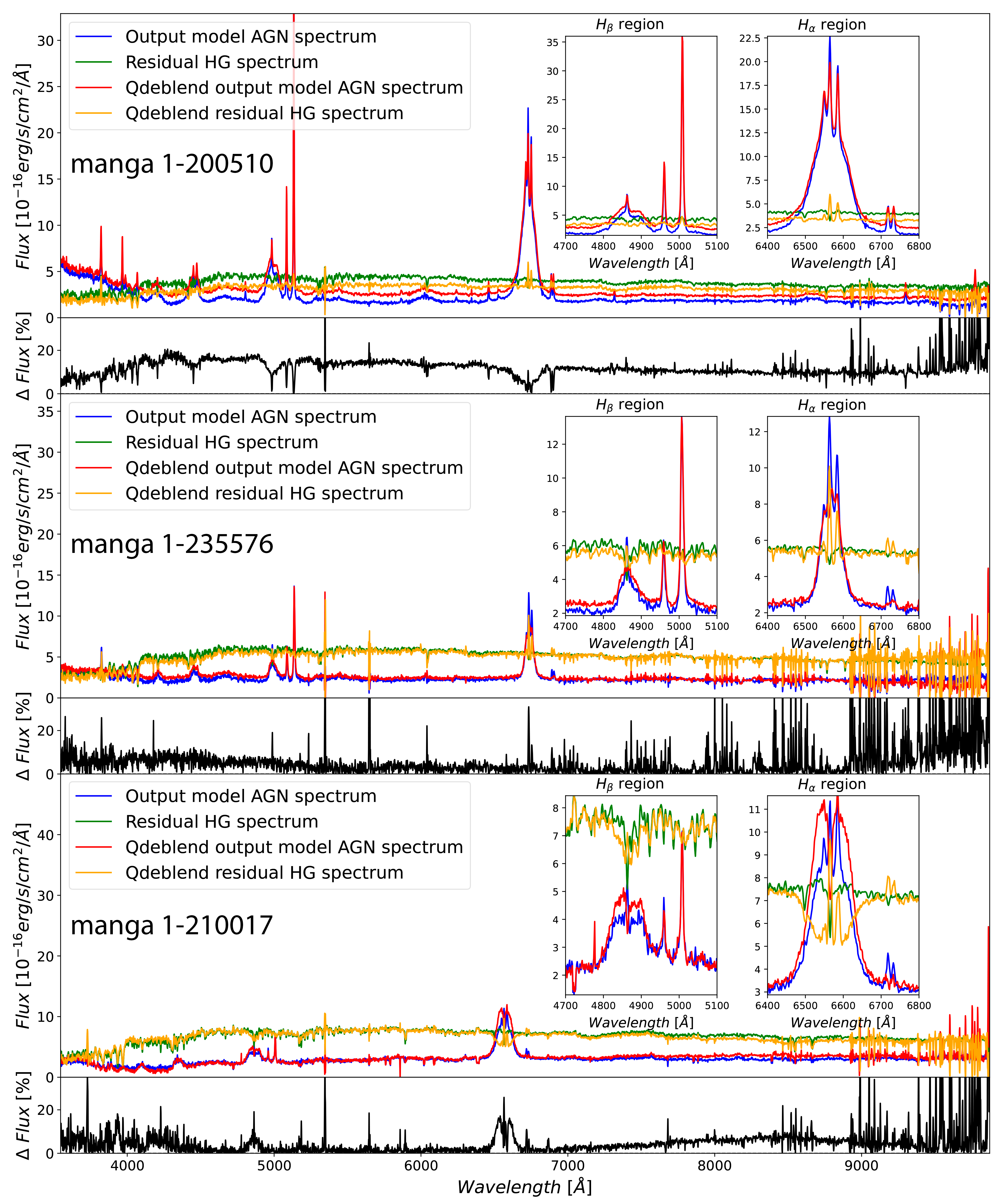}
\caption{{\sc QDeblend3D} comparison of our deblended quasar spectra for MaNGA galaxies 1-200510 (top), 1-235576 (middle), and 1-210017 (bottom). The solid blue line represents the deblended quasar spectra obtained with our methodology, and the solid red line represents the quasar spectra obtained with {\sc QDeblend3D}. The green and yellow lines correspond to the residual of the HG spectrum from our method and {\sc QDeblend3D}, respectively. The two inset plots show a zoom-in of the $H_{\beta}$ and $H_{\alpha}$ regions. The lower panels show the normalised residuals: {\sc QDeblend3D} minus our method over the input flux. The spectrum is the total spectrum within 2.5 arcsecs of the centre of the IFU-FoV.}
\label{fig:qdblend}
\end{figure*}

\section{Implementation on MaNGA data}\label{sec:realdata}

In this Section, we applied our methodology to the MaNGA galaxies 1-200510 (pateid-ifuid: manga-11944-12704, iau: J161301.62+371714.9), 1-235576 (manga-8326-6102, J142004.29+470716.8), and 1-210017 (manga-8549-12702, J160505.14+452634.7). These galaxies host a Type I AGN that was selected with the methodology presented in \citet{Cortes+22} and \citet{Hernandez-Toledo+2023}. The selected galaxies were previously studied in several works \citep[e.g.,][]{Rembold+2017,Sanchez+2019,Wylezalek+2018,Wylezalek+2020,Comerford+2020,Cortes+22,Negus+2023,Hernandez-Toledo+2023} and can be downloaded with the online SDSS Marvin tool\footnote{\url{https://dr17.sdss.org/marvin/}}. In most of those works, they analyze the spectra with the official MaNGA data analysis pipeline \citep[DAP,][]{Westfall+2019}, and with {\sc pyPipe3D} \citep{Sanchez+2022} and the {\sc starlight} tool \citep{Cid-Fernandes+2005} in the case of \citet{Cortes+22} to decouple the HG emission from the AGN spectroscopically. All the works analyze the HG and the AGN spectroscopically by fitting and modelling the emission lines and continuum of the resolved spectra spaxel by spaxel or the integrated spectra within a well-defined aperture.

In Figure \ref{fig:spectral_debl}, we show the deblended AGN and HG spectra for the three galaxies. The deblending code can recover the HG stellar {C\sc{II}} HK (3969.59 and 3934.78 \AA) absorption lines in the residual spectra and recover the line profiles of the AGN's broad and narrow emission lines. In Figure \ref{fig:map_example}, we show the surface brightness maps of \ha. The results show the initial flux distribution, the AGN point-source flux model, and the residual HG emission. In all the cases, our deblending method can model and subtract the AGN emission without generating an over-subtraction or artefacts on the residuals.  

\begin{figure*}
\includegraphics[width=1.9\columnwidth]{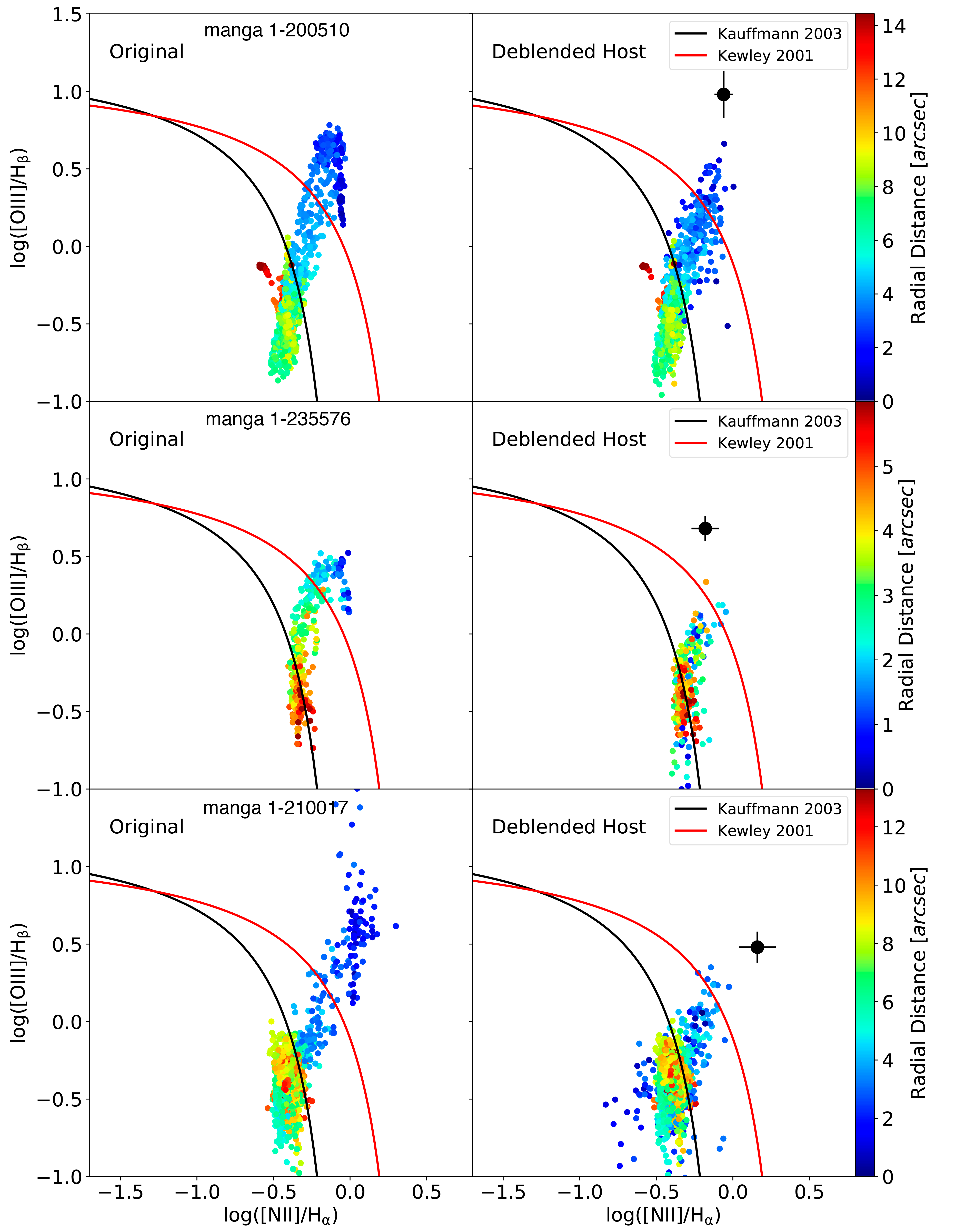}
\caption{The Baldwin, Philips \& Terlevich NII diagram for our three MaNGA galaxies: 1-200510 (top), 1-235576 (middle), and 1-210017 (bottom). The left panels show the spaxel positions on the BPT for the non-deblended galaxy spectra. The right panels show the spaxel's positions on the BPT for the HG deblended spectra. The colour bar represents the radial distance from the IFU central position. The solid red line represents the \citet{Kewley+01} demarcation line. The solid black line represents the \citet{Kauffmann+2003May} delimitation line. The solid black point represents the position of the narrow lines of the deblended quasar spectra. We plot only spaxels with an SNR greater than 5.}
\label{fig:bpt}
\end{figure*}

\subsection{Comparison with QDeblend3D}

Similarly, as in Section \ref{qdb1}, we use the previous three MaNGA galaxies and compare our deblending method with the {\sc QDeblend3D} tool; see Figure \ref{fig:qdblend}. As we discuss in Section \ref{qdb1}, we decide to use only the {\sc QDeblend3D} tool because it does not require modelling the PSF with a foreground star previously, and, therefore, it is the most similar tool to our method. We measure the normalised residual (by the original spectra) between the deblended AGN spectrum obtained with our method and with {\sc QDeblend3D}. 

For the case of 1-200510, which is a high-resolution MaNGA IFS, we obtain a residual of $\approx 20\%$ below 4,000 $\angstrom$, and then it drops to $5\%$ at 3,500 $\angstrom$. The residual also drops at the AGN BELs, which is a consequence of the flux normalisation. In the inset panels of Figure \ref{fig:qdblend}, for 1-200510, we show that both methods are able to recover the same profile for AGN BELs. When the AGN flux is comparable to the HG flux, it is important to note that $20\%$ of the residual of the flux is within the measured $20\%$ precision of our method.  

For the case of 1-235576, which is an intermediate resolution MaNGA IFS, we achieve a good concordance with the AGN deblending of {\sc QDeblend3D}. The BELs are also in good agreement. We note that only the AGN narrow component (NC) is not well recovered for the case of {\sc QDeblend3D}, as we show in the insets of 1-235576 in Figure \ref{fig:qdblend}. 

For the case of 1-210017, a high-resolution MaNGA IFS, we obtain a similar result to that obtained in 1-235576, with the exception of the BEL spectral region. We obtain a good agreement with {\sc QDeblend3D} for the AGN continuum, but our method differs in how the AGN NC of \ha\ and \hb\ are recovered.  From the inset of Figure \ref{fig:qdblend}, the AGN NCs recuperated from {\sc QDeblend3D} show an intense absorption line in the loci of \ha\ where it should be an emission line. In addition, there is an obvious difference between the BEL profiles from the one recovered by {\sc QDeblend3D} and the one retrieved by our method: The BEL from {\sc QDeblend3D} have a larger amplitude. This behavior can be explained by two factors. First, the NCs are under-estimated by {\sc QDeblend3D}, associating the flux of the AGN NCs as part of the HG flux. Second, part of the HG flux, which in this case comes from the galaxy bulge, is associated as part of the AGN BEL emission, then the total BEL emission is overestimated. This overestimation is clearly seen on the HG blended spectra from {\sc QDeblend3D}. This behavior is not a failure of {\sc QDeblend3D}, but it is a consequence of providing a not perfect AGN/HG flux factor as an input value\footnote{see the manual in \url{https://github.com/brandherd/QDeblend3D}}. However, we emphasise that the final performance of the two methods will also depend on the AGN and HG spectrum, for example. For the comparison of the mock cases in Section \ref{qdb1}, we see that {\sc QDeblend3D} returns a lower residual for the AGN NCs.



\begin{figure*}
\includegraphics[width=1\columnwidth]{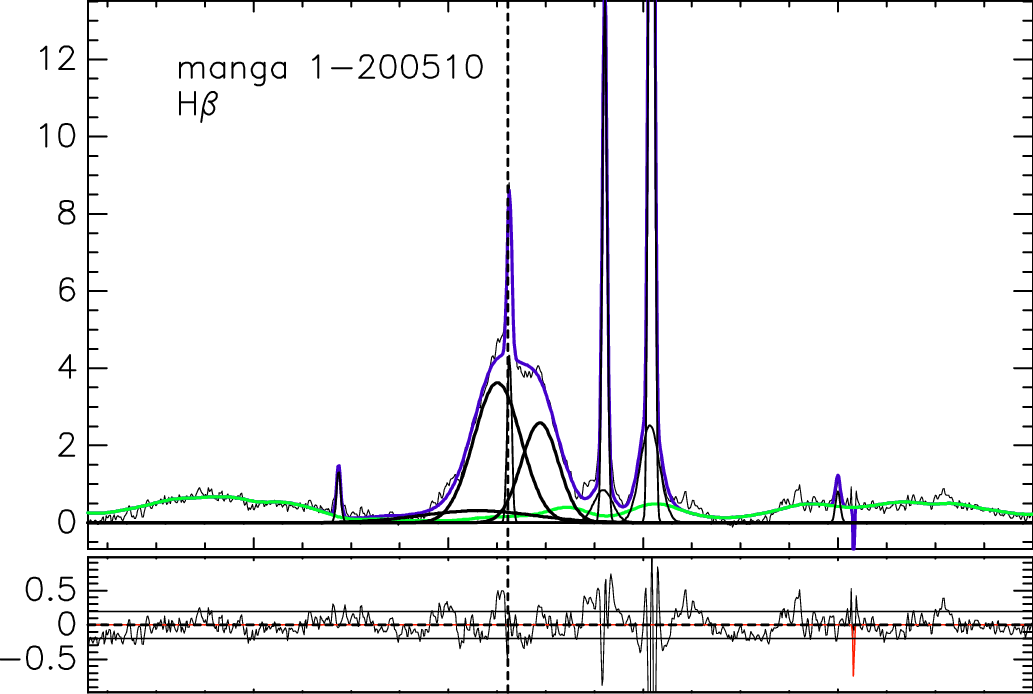}
\includegraphics[width=1\columnwidth]{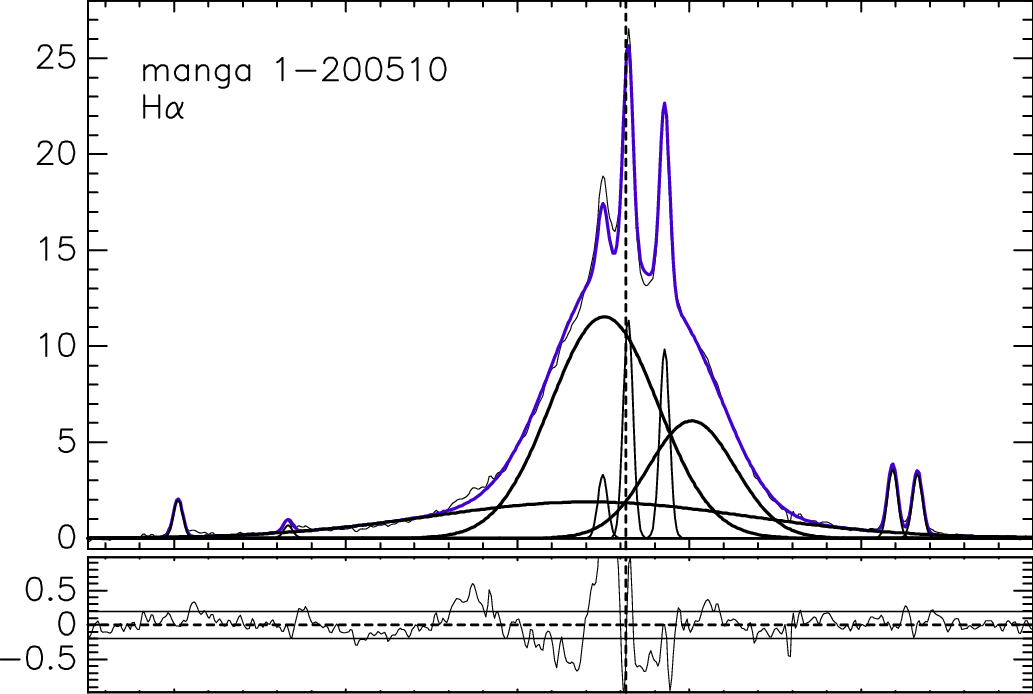}
\includegraphics[width=1\columnwidth]{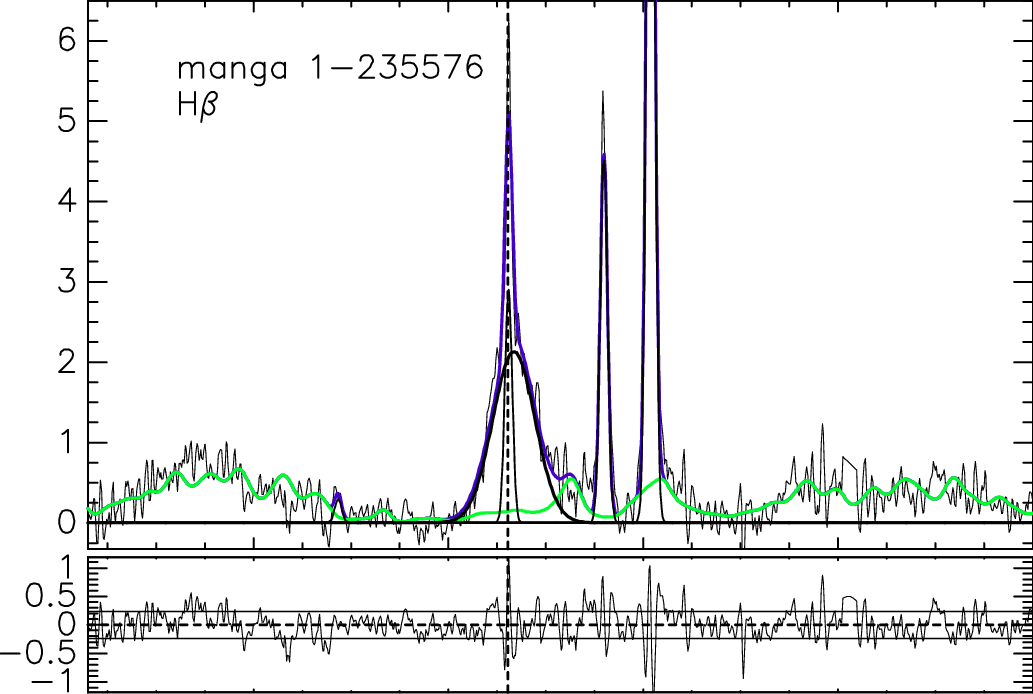}
\includegraphics[width=1\columnwidth]{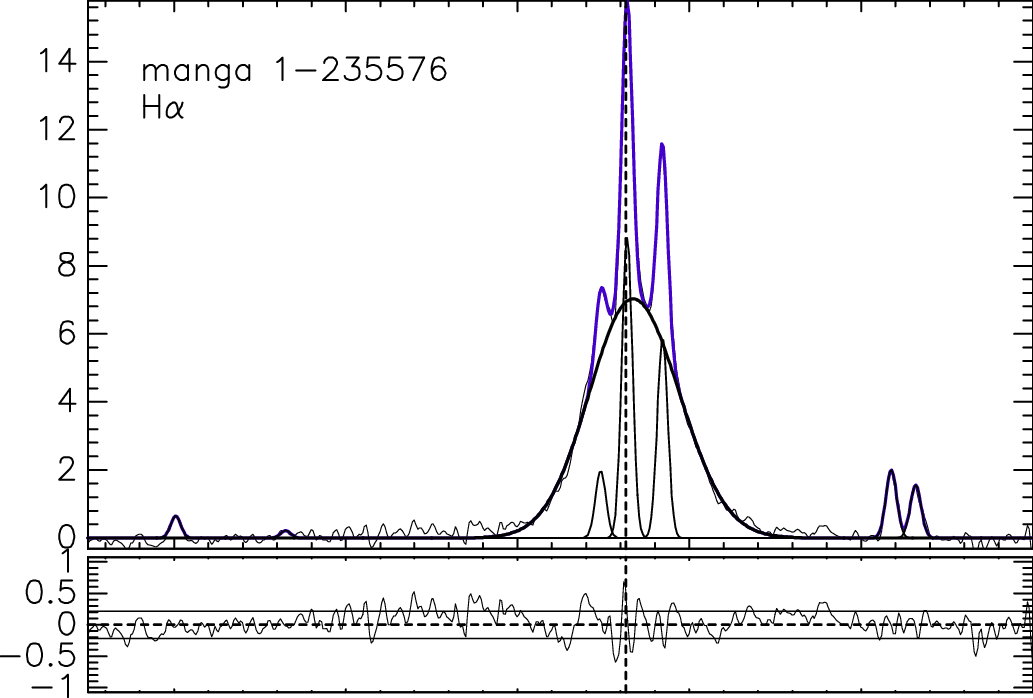}
\includegraphics[width=1\columnwidth]{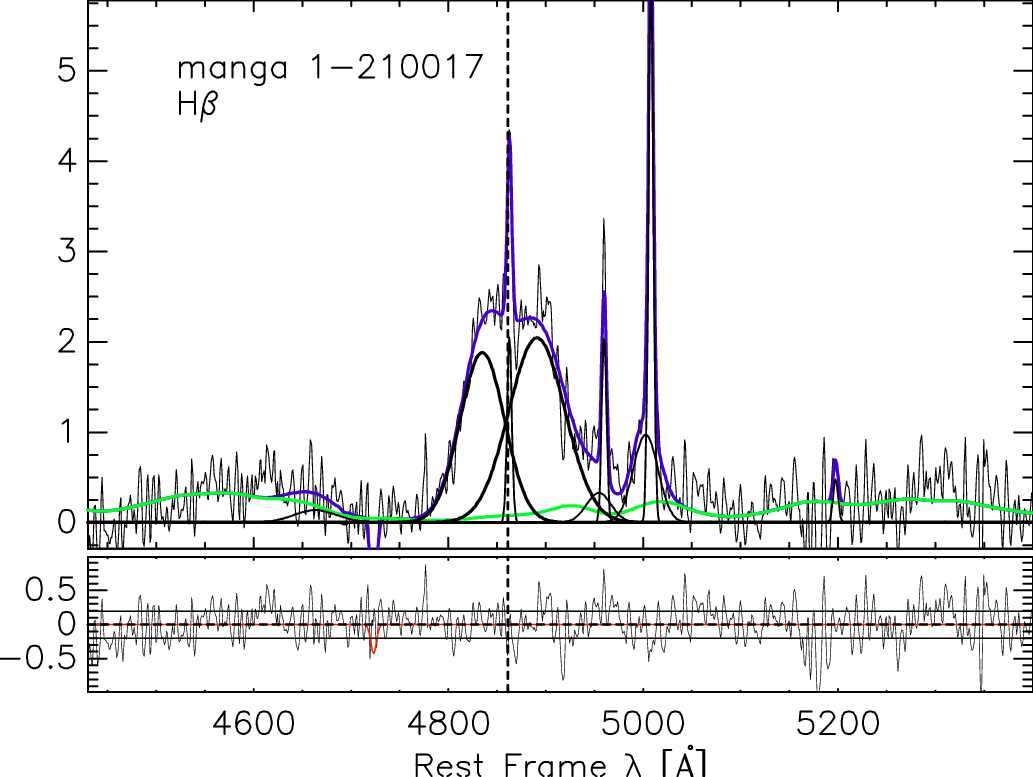}
\includegraphics[width=1\columnwidth]{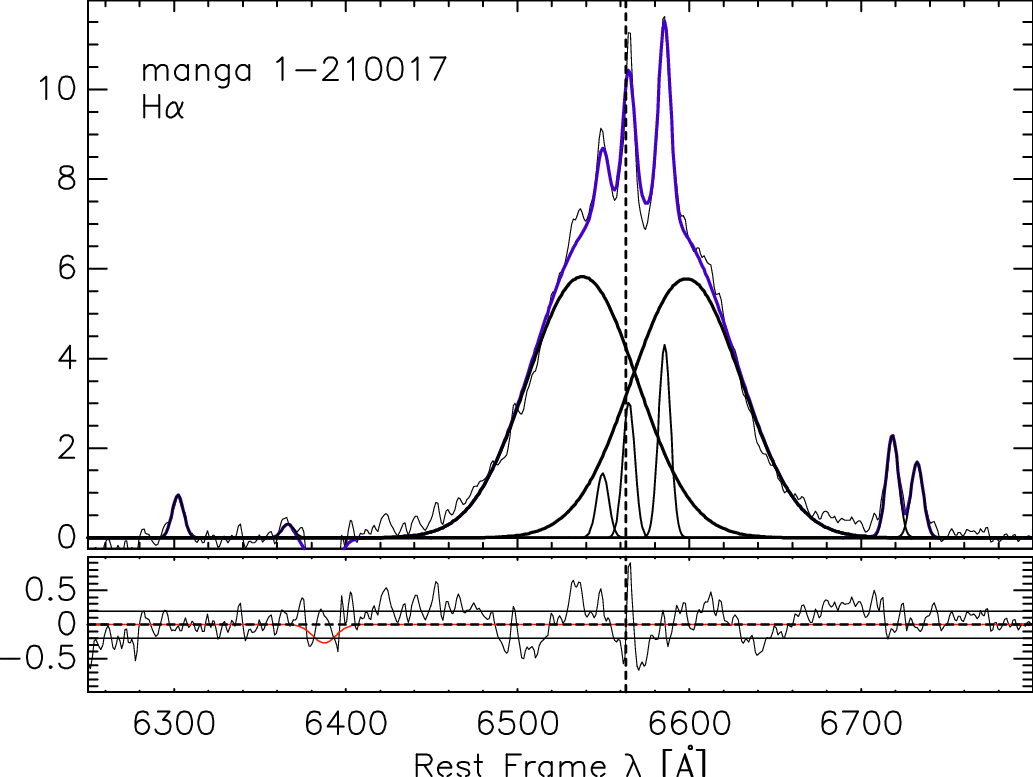}
\caption{{\sc specfit} modelling from the deblended PSF quasar spectra for 1-200510 (top), 1-235576 (middle) and 1-210017 (bottom). For each object, \hb\ spectral region is shown in the left panels and \ha\ in the right panels with a flux scale $\times$ 10$^{-16}$ ergs s$^{-1}$ cm$^{-2}$ \AA$^{-1}$.
The solid black line is the observed spectra, thick black lines are the broad components, and thin black lines represent the narrow components. The purple line is the fit. In the \hb\ region, the green line is the {Fe\sc{ii}} emission. Vertical dotted lines represent the rest-frame wavelength of \hb\ and \ha, respectively. The panels below the fits are the residuals. The solid horizontal line shows the zero level; dashed lines were set at $\pm$ 1$\sigma$ estimated in the continuum window around 5100 \AA. The models of any absorption lines are shown in red.
}
\label{fig:specfit}
\end{figure*}

\subsection{Deblending of the host emission lines: The BPT}

Similarly, as in Section \ref{Mock_BPT}, we explore the deblending of the HG EELRs returned by the method by exploring their BPT diagrams. For all the galaxies, we compare how the BPT diagrams change with and without the AGN/HG deblending process. We use the outputs of the emission line fitting from {\sc pyPipe3D} and generate the BPTs before and after the deblending process; see Figure \ref{fig:bpt}.

The BPT diagrams from the non-deblended galaxy IFUs show that the narrow line ratios of the central spaxels (two arcseconds) have line ratios characteristic of AGNs above the \citet{Kewley+01} line. On the other hand, the line ratios from the deblend HG spectra of the same galaxies show different properties. The central spaxels move mostly from above to below the Kewley line. For the case of the MaNGA galaxy 1-200510 and 1-235576, the average central spaxels are now within the composite region. Finally, for 1-210017, we find the most dramatic change between the unblended/deblended BPT diagrams. For the unblended case, the central spaxels are clearly within the AGN areas, whereas for the host deblended case, the central spaxels are between the composite and the star formation areas. 

\subsection{Deblending of the AGN emission lines}

With the pure AGN deblending spectra, we now proceed to model the \hb (blue) and \ha (red) spectral regions. We use the spectral line fitting from the {\sc IRAF-specfit} routine \citep{kriss94}. The {\sc specfit} routine has the advantage that it can model the underlying continuum and the emission lines at the same time. 
The spectral regions were delimited as follows. The blue one within 4,500-5,400 $\angstrom$ includes the emission lines \hb, \oiiiopt, the \feii\ template, \heiiopt, and \ni. The red one within 6,200-6,800 $\angstrom$ includes \ha, \niill, \siill, and \oill. For all permitted lines, we consider both broad and narrow components. In the case of \oiiiopt, in two objects, it was necessary to add a second semi-broad component. As \oiiiopt\ is a high-ionisation line, this second component has been associated with outflows \citep[e.g.][]{negrete2018}. In both regions, we fit the non-thermal continuum using a nearly flat power law. For all emission lines (except for \feii), we use Gaussian profiles with the following input parameters: a central lambda, the line intensity and the FWHM. We fit the \feii\ emission around \hb\ considering the template by \cite{marziani2009}, which is based on the spectrum of I Zw 1 \citep[see also][]{negrete2018}. {\sc specfit} finds the best fit to the model by minimising the $\chi^2$ using a Marquardt algorithm with 5-10 iterations. In Figure \ref{fig:specfit}, we show the blue and red spectral regions and their models.
 

We started the fit in the red region, starting with the strong broad lines and then adding the narrow ones. For \oiiiopt\ (both narrow and semi broad), \niill\ and \oill, we consider the theoretical ratio 3:1 for its line fluxes \citep{Osterbrock+2006}. We assume that all narrow components come from the same region and thus have the same FWHM shift. Then, we used the \ha\ model as the basis to model \hb. In addition, we model a possible absorption line component to double-check if there is any HG residual.


For the case of the MaNGA 1-200510 galaxy, the fitting of the deblended AGN BLR returns a double broad component for both Balmer lines, one blue-shifted and the other red-shifted, with respect to the systemic restframe. Looking at the residuals, we saw it is necessary to add an additional broader blue-shifted component, much fainter than the first ones. The FWHM of these three broad components are 3,400$\pm$100 and 3,500$\pm$700 \kms\ for the blue components; 2,700$\pm$90 and 2,740$\pm$540 \kms\ for the red components; and 9,950$\pm$700 and 8,500$\pm$300 \kms\ for the broader blue-shifted component in \ha\ and \hb, respectively. The FWHM of the narrow components is 300$\pm$30 \kms\ for both \ha\ and \hb\ regions. The FWHM of the \oiiiopt\ semi-broad component is 1400$\pm$200 \kms. On the other hand, for MaNGA 1-235576, we find that the BLR can be modelled with a single broad component with FWHM of 2,700$\pm$100 and 3,100$\pm$100 \kms, plus a narrow component with FWHM of 300$\pm$30 and 500$\pm$50 \kms, for \ha\ and \hb, respectively. The best fit for the MaNGA 1-210017 galaxy is similar to 1-200510. It returns a double broad component. We find FWHM values of 3,400$\pm$100 and 3,300$\pm$200 \kms\ for the blue components, and 3400$\pm$100 and 4200$\pm$500 \kms\ for the red components, for \ha\ and \hb, respectively. The FWHM of all narrow components is 400$\pm$40 \kms. It is important to note the similarity between \hb\ and \ha\ broad and narrow emission lines for the three cases.

The physical interpretation of the modelled double broad components can be associated with the signature of the kinematics of the accretion disk as reported by, e.g., \citet[][]{Storchi-Bergmann2017}. The double broad components with FWHM up to a few tens of thousands \kms\ are common in low-luminosity AGN and radio galaxies, usually at low-$z$, and can be easily recognized by the top flatted broad component line profile or, in some cases, both broad lines are visually separated. The suggested scenario is the continuity between the inner part of the broad-line region (BLR) and the outer part of the accretion disk. In the case of low-luminosity objects and objects with low Eddington ratio ($L_{bol}/L_{Edd} < 0.1$), such as radio galaxies, the outer material from the BLR can reach the inner part of the central machine, i.e., the accretion disk \citep[e.g.,][]{Eracleous1994,Netzer2013,Popovic2014}. In addition, double BELs with different characteristics and asymmetries for AGNs have also been explained by the residual HG absorption lines contribution \citep[e.g.,][]{Bon2020}.

For the narrow components of the deblended AGN, we obtain the values of \niib/\ha\ and \oiiia/\hb\ line ratios. The measured line ratios indicate a clear ionisation AGN source in the BPT diagram. In Figure \ref{fig:bpt}, we plot the positions of the line ratios on the deblended BPT diagrams. Finally, we find that the modelled absorption lines are comparable with the dispersion of the residuals. Therefore, there is no evidence of the HG absorption features present in the deblended pure AGN spectra.

\section{Summary}\label{sec:summary}

This work presents a new alternative iterative method to perform a spatially resolved AGN deblending for IFS data. This method models the non-resolved AGN emission and the HG-resolved emission simultaneously in each monochromatic slide,  being able to interpolate the spectral variability of the PSF across the wavelength range. This method is ideal for decoupling the intense emission from AGNs when the nuclear flux domains over the HG central region. We summarise the principal characteristics, results, and tests of the method: 

\begin{itemize}
    \item  The resolved/unresolved flux deblending: We explore the precision of the deblended algorithm by implementing a set of six mock IFS data cubes (three resolutions $\times$ two AGN dominances). Due to the two-dimensional surface brightness modelling, the HG deblending precision on the EELR depends on the effective spatial resolution of the IFS and the PSF size along the IFU FoV (Figs. \ref{fig:high_resolution}--\ref{fig:low_resolution}). The better the PSF is resolved, the better the method disentangles the narrow emission lines from the HG. In addition, the method performs better when the AGN flux is more intense. 
    \item Our method uses the S\'ersic HG modelling as an initial background value to model the non-resolved AGN emission as a PSF profile. Therefore, the morphologies have little impact on the performance of our deblending method. For the mock cases, the HighR and IntR mocks, the input HGs present spiral and bar structures that can not be modelled with a simple S\'ersic profile. On the contrary, the LowR mock has a morphology that can be modelled with the S\'ersic profile. Still, the best decouple performance is for the HighR, which supports that the morphology has less impact on the deblending process.
    \item The method can perform a good deblending of the AGN and HG spectral continuum within $20\%$ of the normalised residual flux.  In addition, we performed a detailed analysis of the deblending of the HG EELR and the AGN BEL and NC (Figs. \ref{fig:ELMock} and \ref{fig:ELMockE}). The central HG EELR fluxes are deblended within $13\%$, $30\%$ and $40\%$ for the HighR, IntR and LowR spatial resolution cases, showing that our method is sensitive to the IFS spatial resolution. For the EELR velocity shifts, we recuperate the true central values within $30\%$ without an explicit dependency on the spatial resolution or the flux ratio between the AGN and the HG. In addition, for the case of the recovered FWHM of EELRs, the method can recuperate the true values within $15\%$. On the other hand, for the case of the AGN BEL and NC components, the method performs very well recuperating the true values within the $9\%$.
    \item The Deblended HG SSPs: With the mock IFS, we explore how well the methodology disentangles the HG spectra to estimate the SSP $Age_{LW}$. The SSP synthesis recovers the values of the central age once the estimation errors are considered. However, the central Age values tend to be slightly younger, and at larger radii, the Ages tend to be older but always within the error estimations (Fig. \ref{fig:ages}) Indeed, we need to be cautious in the interpretation of the SSP results due to any possible bias that could be propagated by the uncertainties of the AGH/HG decomposition.
    \item Comparison with {\sc QDeblend3D}: We compare our deblending method with the well-tested two-dimensional quasar deblending {\sc QDeblend3D} tool. From the mock analysis, we find that both methods achieve a similar precision above 5000 \AA. However, {\sc QDeblend3D} tends to produce a bluer AGN spectrum (or a redder HG spectrum) below 5000 \AA\ and maintains a stable residual above 7000 \AA. In contrast, our method tends to produce lower residuals across the full spectral range. On the other hand, {\sc QDeblend3D} recovers almost the same AGN continuum as that deblended by our method. For the case where we did not have a perfect match with the output of {\sc QDeblend3D}, we still have a good agreement with the AGN spectral continuum, with similar values of the normalised residuals (Fig. \ref{fig:qdblendMock}). Both tools perform similarly for emission lines, although {\sc QDeblend3D} tends to overestimate the BEL regions, while depending on the object, the residuals of the NC regions tends to be better in {\sc QDeblend3D}. Overall, the performance of both methods varies with spectral characteristics, likely due to limitations in flux input data, with each method returning a better decomposition on the NC regions depending on the object (Fig. \ref{fig:qdblend}).
    \item The deblended HG EELRs from the BPT diagram: We measure the HG ionisation properties of the resolved emission lines of the spectra before and after applying our AGN/HG deblending method. We use the mock IFS observation and three real MaNGA data with ids 1-200510, 1-235576, and 1-210017. We found that the central emission line rations move on the NII-BPT diagram from the AGN ionisation region (before deblending) to the compose/SFR region (after deblending). Therefore, the deblending method can effectively disentangle the AGN emission from the HG emission. 
    \item The deblended AGN emission lines profiles: With the three MaNGA cubes, we test the quality of the deblended AGN broad and narrow emission lines with a detailed fitting of the line profiles. We perform individual line modelling at the \hb\ and \ha\ complex and find that both regions can be fitted with a similar model. These models show that the BELs have a well-defined profile at each spectral region. This profile similarity is expected from a well-deblended AGN spectrum. In addition, we measure the position of the emission line ratios on the NII BPT diagram, confirming that the narrow lines have a clear AGN ionisation origin.
\end{itemize}

For all previous tests, we show that the method is able to disentangle the intense central emission of their central AGN in IFS data. These tests show that our method can decouple the narrow nebular emission of the HG from the AGN emission and its stellar continuum, and provide pure host-free total quasar spectra. The advantage of this method is that it does not require any previous information on the PSF or HG SB profiles. The method models the PSF using the strong central AGN emission from the input IFS observation. However, this principle also came with a drawback: the model depends on how much information about the PSF can be retrieved from the IFS observation; how much the PSF is resolved within the IFS and how much it is the contrast of the non-resolved emission in comparison with the resolved HG emission. Therefore, the method depends on the spatial resolution of the PSF within the IFU FoV and the relative flux intensity of the AGN. In addition, there are inaccuracies with the HG 2D SB profile model, a drawback that we plan to attack in a future release of the method by implementing a more complex HG modelling that includes bars, discs and different asymmetries.

\section*{Acknowledgements}
HIM acknowledges the support from grant CONAHCyT CBF2023-2024-1418, IN-106823 PAPIIT UNAM, IN-119123 PAPIT UNAM and CONHACyT CF-2023-G-543. CAN acknowledge support from grant IN111422 PAPIIT UNAM, and CONAHCyT project Paradigmas y controversias de la Ciencia 2022-320020. HMHT acknowledges support from grants  CONAHCYT project CF-G-543 entitled “Arqueolog\'ia
y filogen\'etica de dinosaurios gal\'acticos: formaci\'on y evoluci\'on de galaxias masivas apagadas” and CONAHCYT project CF-2023-G-1052 entitle “Sinergia y Retos del Censo LSST del Observatorio Vera Rubin para la Astrof\'isica, la Ciencia de datos, la Qu\'imica y otras disciplinas”. ECS thanks CONAHCyT for the academic postdoctoral stay scholarship (CVU 825458). This paper made use of the pyPipe3D code. The authors thank the IA-UNAM MaNGA team for creating this tool.

We acknowledge the SDSS-IV collaboration for making publicly available the data used in this paper. SDSS-IV is managed by the Astrophysical Research Consortium for the Participating Institutions of the SDSS Collaboration, including the Brazilian Participation Group, the Carnegie Institution for Science, Carnegie Mellon University, the Chilean Participation Group, the French Participation Group, Harvard-Smithsonian Center for Astrophysics, Instituto de Astrof\'isica de Canarias, The Johns Hopkins University, Kavli Institute for the Physics and Mathematics of the Universe (IPMU)/University of Tokyo, Lawrence Berkeley National Laboratory, Leibniz Institut f\"ur Astrophysik Potsdam (AIP), Max-Planck-Institut f\"ur Astronomie (MPIA Heidelberg), Max-Planck-Institut f\"ur Astrophysik (MPA Garching), Max-Planck-Institut f\"ur Extraterrestrische Physik (MPE), National Astronomical Observatories of China, New Mexico State University, New York University, University of Notre Dame, Observat\'ario Nacional / MCTI, The Ohio State University, Pennsylvania State University, Shanghai Astronomical Observatory, United Kingdom Participation Group, Universidad Nacional Aut\'onoma de M\'exico, University of Arizona, University of Colorado Boulder, University of Oxford, University of Portsmouth, University of Utah, University of Virginia, University of Washington, University of Wisconsin, Vanderbilt University, and Yale University.

\section*{Data availability}
The methodology underlying this article is publicly available in the GitHub repository located at \url{https://github.com/hjibarram/AGN_decompose}. The data underlying this article are available in the MaNGA datasets from the public domain using the SDSS-IV MaNGA Public Data Release 17, at \url{https://www.sdss.org/dr17/}. 

\appendix

\bibliographystyle{mnras}
\bibliography{example} 

\bsp	
\label{lastpage}
\end{document}